%% file: main.tex
\documentclass[preprint,12pt]{elsarticle}

\usepackage{amsmath,amssymb,amsfonts}
\usepackage{color}
\usepackage{mathrsfs}
\usepackage{algorithm}
\usepackage{algpseudocode}
\usepackage{booktabs}
\usepackage{array}
\usepackage{float}
\usepackage{graphicx}
\usepackage{subcaption}
\usepackage{lineno}
\usepackage{hyperref}

\journal{Journal of Computational Physics}

\newcommand{\bu}{\boldsymbol{u}}

\newcommand{\bn}{\boldsymbol{n}}
\newcommand{\bfv}{\boldsymbol{f}}
\newcommand{\bomega}{\boldsymbol{\omega}}
\newcommand{\btau}{\boldsymbol{\tau}}
\newcommand{\bA}{\boldsymbol{A}}
\newcommand{\bI}{\boldsymbol{I}}
\newcommand{\bx}{\boldsymbol{x}}
\newcommand{\bz}{\boldsymbol{z}}
\newcommand{\bq}{\boldsymbol{q}}
\newcommand{\cL}{\mathcal{L}}
\newcommand{\cH}{\mathcal{H}}
\newcommand{\cM}{\mathcal{M}}
\newcommand{\cQ}{\mathcal{Q}}
\newcommand{\cD}{\mathcal{D}}
\newcommand{\IG}{\ensuremath{\mathrm{IG}}}

\begin{document}
\nolinenumbers

\begin{frontmatter}

\title{Invariant Guided PINN for Fluid Flow Computation}

\author[addr1]{Zheng Lu}
\ead{luzheng21@mails.jlu.edu.cn}
\author[addr1,addr3]{Jiwei Jia\corref{cor1}}
\ead{jiajiwei@jlu.edu.cn}
\author[addr2]{Bora Aniruddha}
\ead{qvx13@txstate.edu}
\author[addr2]{Xingyu An}
\ead{mbo42@txstate.edu}
\author[addr2]{Young Ju Lee\corref{cor2}}
\ead{yjlee@txstate.edu}

\cortext[cor1]{Corresponding author.}
\cortext[cor2]{Corresponding author.}

\address[addr1]{School of Mathematics, Jilin University, Changchun, Jilin, 130012, China}
\address[addr2]{Department of Mathematics, Texas State University, San Marcos, TX, 78666, USA}
\address[addr3]{AI for Science and Engineering Center, Shenzhen Loop Area Institute, Shenzhen 518048, China}

\begin{highlights}
\item Partitioned PINN training is used as conservative preconditioning.
\item Global correction yields one full-domain neural flow representation.
\item Oldroyd--B transfer uses velocity, stress, and mass-flux information.
\item Hard-time helicity transfer controls energy and structure drift.
\end{highlights}

\begin{abstract}
Physics-informed neural networks (PINNs) often become difficult to optimize for
incompressible flow problems with large spatial domains, multiscale stresses, or
long-time invariant dynamics. We propose an invariant-guided PINN
(IG-PINN) framework that uses partitioned training as a conservative
preconditioning stage rather than as the final piecewise representation. A
globally defined architecture is trained successively on spatial subdomains or
temporal slabs; selected field traces, structural information, and conservative
diagnostics are then transferred to a final global correction, yielding a single
neural field on the full spatial or space-time domain. The framework is tested
on two incompressible flow problems: steady Oldroyd--B flow past a confined
cylinder and a rotational Newtonian flow with helicity diagnostics. In the
Oldroyd--B case, IG-PINN transfers velocity, polymeric stress, and mass-flux
information while avoiding pressure traces at artificial interfaces. In the
helicity case, endpoint velocity is transferred through a hard temporal
constraint and kinetic energy is controlled during slab training and residual
global correction. The experiments demonstrate improved optimization robustness,
reduced conservation errors for the cylinder wake, and controlled energy and
helicity diagnostics for the transient rotational flow.
\end{abstract}


\end{frontmatter}

\input{sections/introduction}
\input{sections/governing_equations}
\input{sections/method}
\input{sections/numerical_experiments}
\input{sections/discussion}
\input{sections/conclusion}
\input{sections/declarations}
\input{sections/acknowledgements}

\bibliographystyle{elsarticle-num}
\bibliography{references}

\end{document}

%% file: sections/introduction.tex
\section{Introduction}
\label{sec:introduction}

Physics-informed neural networks (PINNs) have become a widely used framework
for solving forward and inverse problems governed by partial differential
equations \cite{raissi2019physics}. A PINN represents the unknown solution by a
neural network and trains the network by minimizing residuals of the governing
equations together with boundary and initial conditions. This strong-form
construction is attractive for incompressible flow problems because automatic
differentiation gives direct access to the differential operators appearing in
momentum balance, divergence constraints, stress transport, vorticity, and
pressure gradients \cite{baydin2018automatic}. PINN-type models have therefore
been developed for incompressible Navier--Stokes flows, surrogate flow
prediction, and Reynolds-averaged turbulence closures
\cite{jin2021nsfnets,sun2020surrogate,rao2020physics,eivazi2022rans}.
However, direct global PINN training often becomes unreliable when the spatial
domain is large, the time interval is long, or the flow contains strongly
localized structures. The resulting optimization problem
may be stiff, different loss terms may compete across widely separated regions,
and the spectral bias of neural networks may delay the resolution of localized
high-gradient features
\cite{rahaman2019spectralbias,wang2022when}. In practice, these difficulties
can lead to slow convergence, nearly trivial local minimizers, and loss of
physically important conservation or invariant properties.

These issues are particularly severe for complex incompressible flow problems.
In non-Newtonian flows, the velocity field is coupled to nonlinear stress or
conformation dynamics, and small errors in the velocity gradient may be
amplified through the constitutive equation. This difficulty is closely related
to the high-Weissenberg-number challenge in viscoelastic computation and to the
need for positive-definite or logarithmic conformation representations
\cite{baaijens1998mixed,owens2002computational,fattal2005time,
lozinski2003energy,balci2011symmetric,castillo2022understanding}. Recent
data-driven and PINN-based studies have begun to address non-Newtonian fluid
models, but robust simulation without reference data remains difficult for
large confined geometries
\cite{reyes2021learning,mahmoudabadbozchelou2022nn,nguyen2022physics,
kumar2021physics}. In long-time rotational flows, small errors in velocity,
vorticity, or divergence may accumulate and produce artificial drift in
geometric or energetic quantities such as helicity and kinetic energy. Thus,
the main challenge is not only to reduce the residual of the governing
equations pointwise, but also to preserve the physically relevant information
that should be transported across space or time.

Domain decomposition PINNs address part of this difficulty by replacing one
large optimization problem with smaller subproblems. Conservative PINNs
(cPINNs), extended PINNs (XPINNs), parallel PINNs, and finite-basis PINNs
(FBPINNs) demonstrate that decomposing the computational domain can improve
scalability and local representation power
\cite{jagtap2020conservative,jagtap2020extended,shukla2021parallel,Moseley2023}.
Nevertheless, decomposition alone does not guarantee that the information
learned locally is effectively transferred to a globally consistent solution.
Interface penalties may over-constrain quantities that are not unique, most
notably pressure in incompressible flow; weak interface matching may leave
visible discontinuities or gradient inconsistencies; and a collection of
accurate local networks does not by itself provide a single globally corrected
neural representation. Schwarz-type neural domain decomposition methods offer a
different route by repeatedly solving subdomain problems and exchanging
boundary data, but such alternating iterations may be expensive when each
subproblem is itself a neural optimization problem.

We propose an invariant-guided PINN framework, abbreviated
IG-PINN, for complex flow equations with physically meaningful invariant or
balance structures. The key idea is to use partitioned training as a
conservative preconditioning stage rather than as the final piecewise neural
representation. Partition stages are first trained on spatial subdomains or
temporal slabs; in the unified implementation used here, the same global
architecture is continued from one partition stage to the next. From these
stage solutions, we extract physically transferable information, including
selected field traces, conservative diagnostics, and structural admissibility
information. These quantities are then used to initialize or constrain
subsequent training problems and, finally, a global correction model. The final
output is a single neural field defined over the full spatial domain or full
space-time cylinder.

The transfer variables in IG-PINN are selected according to the physical
structure of the equation being solved. They are not obtained by blindly copying
all network outputs across artificial interfaces. For incompressible flow, this
distinction is important because the pressure is determined only up to a gauge
unless an additional normalization is imposed. Passing pressure traces between
subdomains can therefore introduce nonphysical constraints. Instead, the
proposed framework transfers only physically meaningful information, such as
velocity traces, stress traces, mass fluxes, endpoint velocity states, and
energy levels. Structural constraints, such as positive definiteness of the
conformation tensor or curl-compatibility of the vorticity, are incorporated at
the representation level whenever possible. This viewpoint is also consistent
with hard-constraint PINN constructions, where admissible boundary or
structural behavior is built into the neural representation rather than left
entirely to soft penalties
\cite{lu2021hard,sukumar2022exact}.

We test the proposed framework on two representative complex flow problems. The
first is a steady non-Newtonian flow past a confined cylinder governed by the
Oldroyd--B model. This benchmark is challenging for direct global PINN training
because the long channel separates the inflow boundary from the cylinder
region, while the stress equation requires stable treatment of the conformation
tensor. The confined-cylinder Oldroyd--B benchmark has also served as a
standard test case for stabilized finite-volume, spectral, and finite-element
viscoelastic solvers
\cite{dou1999oldroyd,dou2007viscoelastic,claus2013viscoelastic,
alves2001flow,pimenta2016stabilization,wittschieber2022stabilized}. IG-PINN
partitions the domain in the streamwise direction, trains one
shared network successively on the inlet, cylinder, and outlet subdomains,
transfers velocity and polymeric stress but not pressure, enforces mass-flux
consistency along vertical sections, and then continues the same architecture
with a full-domain global correction.

The second problem is a transient Newtonian flow written in rotational form,
where the relevant geometric diagnostic is fluid helicity. In this case, the
framework is local in time but global in space during the slab stage. Each time
slab is trained by a global-in-space PINN warm-started from the previous slab,
the terminal velocity is passed to the next slab through a hard temporal
constraint, and an energy-consistency constraint prevents the temporal sequence
from drifting toward an unphysical state. After all slabs have been trained, the
slab sequence is used as a teacher for a residual global space-time correction
on $[0,T]\times\Omega$. The vorticity is computed as
\[
\bomega_\Theta=\nabla\times\bu_\Theta,
\]
rather than learned as an independent output, thereby preserving the
curl-compatibility relation at the representation level and reducing a common
source of artificial helicity drift. The focus on helicity follows the
classical view of helicity as a geometric and topological flow invariant and
the modern numerical emphasis on structure-preserving discretizations
\cite{berger1984topological,moffatt1981some,moffatt1992helicity,
moffatt2014helicity,liu2004energy,layton2008helicity,rebholz2007energy,
olshanskii2010note,kraus2017variational,hu2021helicity,
gawlik2020conservative}.

The main contributions of this work are summarized as follows.
\begin{itemize}
\item We introduce IG-PINN, an invariant-guided training
framework for complex incompressible flow problems with invariant or balance
structures. The method uses partitioned training as a conservative
preconditioning stage and produces a single globally corrected neural
solution.

\item We develop a spatial invariant-guided transfer strategy for the
Oldroyd--B confined-cylinder problem. The method transfers velocity,
polymeric stress, and mass-flux information while avoiding pressure traces at
artificial interfaces, thereby reducing nonphysical interface constraints in
incompressible viscoelastic flow.

\item We develop a temporal hard-constrained transfer strategy for long-time
rotational Newtonian flow. Endpoint velocity fields are propagated between
time slabs by construction, an energy-consistency constraint stabilizes the
temporal sequence, and a residual global correction reduces helicity and
divergence drift while staying close to the slab trajectory.

\item Numerical experiments show that IG-PINN improves optimization
robustness relative to direct global training, reduces conservation errors in
the Oldroyd--B benchmark, and limits long-time energy and helicity drift in
the rotational flow benchmark.
\end{itemize}

The remainder of the paper is organized as follows.
Section~\ref{sec:governing} introduces the governing equations and the
conservative quantities used in the two model problems.
Section~\ref{sec:method} presents the IG-PINN framework, including local
partition training, physically transferable quantities, and global correction.
Section~\ref{sec:numerics} reports the numerical experiments for the
Oldroyd--B cylinder problem and the rotational Newtonian helicity problem.
Finally, Section~\ref{sec:conclusion} concludes the paper and discusses
possible extensions.

%% file: sections/governing_equations.tex
\section{Model Problems and Transferable Physical Quantities}
\label{sec:governing}

We consider two incompressible flow problems that expose complementary failure
modes of physics-informed neural solvers for complex flow equations. The first
is a steady viscoelastic flow past a confined cylinder. In this problem, the
main difficulty is spatial: the velocity field, pressure, and polymeric stress
are strongly coupled, and local errors in the velocity gradient may be
amplified through the conformation equation. This example is therefore used to
test spatial invariant-guided transfer. The second problem is a three-dimensional
rotational Newtonian flow. In this case, the main difficulty is temporal:
small errors in velocity, divergence, or curl may accumulate over long time
intervals and lead to artificial drift of energy and helicity. This example is
therefore used to test temporal hard transfer and energy-consistent global
correction.

Although the two models have different physical origins, they play the same
algorithmic role in the proposed IG-PINN framework. Each model identifies a
set of physically transferable quantities that can be extracted from a
partitioned training stage and used to constrain a subsequent global
correction. For the Oldroyd--B problem, the transferable information consists
of velocity--stress traces and the steady mass flux. For the rotational
Newtonian problem, the transferable information consists of endpoint velocity
states, an energy-consistency constraint, and a curl-compatible vorticity
representation. These quantities are not arbitrary network outputs. They are
selected from the conservation, balance, or structural properties of the
governing equations.

\begin{table}[t]
\centering
\caption{Algorithmic role of the two model problems in the IG-PINN framework.}
\label{tab:transfer_roles}
\resizebox{\linewidth}{!}{%
\begin{tabular}{lp{0.35\linewidth}p{0.35\linewidth}}
\toprule
Model problem & Dominant difficulty & Transfer mechanism \\
\midrule
Oldroyd--B cylinder flow
& Spatial nonlinearity and stress coupling
& Velocity--stress traces and mass flux \\
Rotational Newtonian flow
& Long-time invariant drift
& Endpoint velocity and energy consistency \\
\bottomrule
\end{tabular}
}
\end{table}

\subsection{Steady Oldroyd--B flow past a cylinder}
\label{subsec:oldroyd_model}

Let \(\Omega_{\mathrm{OB}}\subset\mathbb{R}^2\) be a two-dimensional channel
with an embedded circular cylinder, a benchmark geometry widely used in
computational studies of viscoelastic Oldroyd--B flow
\cite{oldroyd1950formulation,alves2001flow,dou1999oldroyd,owens2002lust,
goyal2012direct}. The boundary is decomposed into the inlet, outlet, channel
walls, and cylinder surface. The unknowns are the velocity $\bu=(u,v)^T$, the
pressure $p$, and the conformation tensor $\bA$. The polymeric extra stress is
recovered from
\[
\btau = \frac{1-\beta_s}{Wi}(\bA-\bI),
\]
where $Wi$ is the Weissenberg number and $\beta_s$ denotes the solvent
viscosity ratio. In conformation form, the dimensionless steady Oldroyd--B
system used in the computations is
\begin{align}
Re(\bu\cdot\nabla)\bu + \nabla p
- \beta_s\Delta\bu - \nabla\cdot\btau
  &= \boldsymbol{0},
  \label{eq:oldroyd_momentum}\\
  \nabla\cdot\bu &= 0,
  \label{eq:oldroyd_incompressibility}\\
  (\bu\cdot\nabla)\bA
  -(\nabla\bu)\bA
  -\bA(\nabla\bu)^T
  +\frac{1}{Wi}(\bA-\bI)
  &= \boldsymbol{0}.
  \label{eq:oldroyd_conformation}
  \end{align}
  The cylinder experiment reported below is carried out in the creeping-flow
  regime. The inertial term is retained in
  \eqref{eq:oldroyd_momentum} only to keep the notation consistent with the
  general dimensionless model.

No-slip boundary conditions are imposed on the channel walls and on the
cylinder surface. At the inlet boundary \(\Gamma_{\mathrm{in}}\), we prescribe
a fully developed inflow velocity \(u_{\mathrm{in}}\) together with the
corresponding fully developed Oldroyd--B inlet stresses. The associated inlet
mass flux is
\[
\cM_0 = \int_{\Gamma_{\mathrm{in}}} \bu_{\mathrm{in}}\cdot\bn_{\mathrm{in}}\,ds ,
\]
with the sign convention chosen so that \(\cM_0>0\) for inflow through the
channel.

For any vertical fluid section
\[
\Gamma_\xi=\{(\xi,y)\in\Omega_{\mathrm{OB}}\},
\]
we define the predicted mass flux by
\[
\cM_\Theta(\xi) = \int_{\Gamma_\xi} u_\Theta(\xi,y)\,dy,
\]
where $u_\Theta$ is the velocity component represented by the neural field.
For an incompressible steady through-flow, $\cM_\Theta(\xi)$ should remain
consistent with the inlet flux. We therefore use the flux discrepancy
\[
\cM_\Theta(\xi)-\cM_0
\]
as the main conservative scalar in the Oldroyd--B experiment. The corresponding
loss is
\begin{equation}
\cL_{\mathrm{flux}}(\Theta) =
\frac{1}{N_f}
\sum_{j=1}^{N_f}
\left|
\cM_\Theta(x_j)-\cM_0
\right|^2 .
\label{eq:flux_loss}
\end{equation}
This term does not replace the pointwise Oldroyd--B residuals. Instead, it
provides a low-dimensional conservative summary of the local neural solution.
This summary is later used in the invariant-guided stage. The idea is analogous in
spirit to enforcing flux or balance information across artificial interfaces in
conservative domain-decomposition methods
\cite{jagtap2020conservative,jagtap2020extended,shukla2021parallel}, but here
the partitioned networks are not intended to define the final piecewise
approximation; they are used to generate physically transferable information
for a single global neural field.

The drag coefficient is evaluated a posteriori from the total stress on the
cylinder,
\begin{equation}
C_D =
\frac{2}{\rho U^2 D}
\int_{\Gamma_c}
\left[
\left(
-p\bI
+\beta_s(\nabla\bu+\nabla\bu^T)
+\btau
\right)\bn
\right]\cdot\boldsymbol{e}_1\,ds,
\label{eq:drag}
\end{equation}
where \(\Gamma_c\) is the cylinder boundary, \(U\) is the characteristic inlet
speed, \(D\) is the cylinder diameter, \(\bn\) is the outward normal to the
cylinder, and \(\boldsymbol{e}_1\) is the streamwise unit vector. The drag is
not used as a training constraint; it serves as an independent integral
diagnostic for comparing the global accuracy of different neural solvers.

\paragraph{Role in IG-PINN.}
For this problem, IG-PINN uses spatial partitioning to reduce the optimization
difficulty caused by the long channel and nonlinear stress coupling. The
transferable field information consists of velocity and polymeric stress
traces. The conservative scalar is the mass flux \(\cM_\Theta(x)\). Pressure
is excluded from the transfer because of its gauge dependence in incompressible
flow. The final correction produces a single full-domain neural field.

\subsection{Rotational Newtonian flow and helicity}
\label{subsec:helicity_model}

The second model problem is a three-dimensional incompressible Newtonian flow
written in rotational form on a bounded spatial domain
\(\Omega_{\mathrm{H}}\subset\mathbb{R}^3\). For notational simplicity, we
write \(\Omega=\Omega_{\mathrm{H}}\) in this subsection.
The unknowns are the velocity \(\bu\), the vorticity \(\bomega\), and the total
pressure
\[
p=\widetilde p+\frac12|\bu|^2 .
\]
The governing equations are written in Lamb rotational form
\cite{lamb1932hydrodynamics}:
\begin{align}
\partial_t\bu
-\bu\times\bomega
+ Re^{-1}\nabla\times\bomega
+ \nabla p
  &= \bfv,
  \label{eq:lamb_momentum}\\
  \bomega &= \nabla\times\bu,
  \label{eq:curl_velocity}\\
  \nabla\cdot\bu &= 0,
  \label{eq:lamb_div}
  \end{align}
  where \(Re\) is the Reynolds number and \(\bfv\) is a prescribed forcing term.
  The boundary conditions are
  \[
  \bu\times\bn=\boldsymbol{0},
  \qquad
  p=0
  \qquad
  \mbox{on } \partial\Omega .
  \]

A central diagnostic for this problem is the fluid helicity,
\begin{equation}
\cH_f(t) = \int_\Omega \bu(t,\bx)\cdot\bomega(t,\bx)\,d\bx .
\label{eq:helicity}
\end{equation}
For smooth solutions satisfying the above boundary conditions, the helicity
balance reads
\begin{equation}
\frac{d}{dt}\cH_f(t) =
-2Re^{-1}\int_\Omega (\nabla\times\bomega)\cdot\bomega\,d\bx
+ 2\int_\Omega \bfv\cdot\bomega\,d\bx .
\label{eq:helicity_identity}
\end{equation}
Thus, in the inviscid and force-free limit, helicity is conserved. At large
Reynolds number, however, a neural time integrator may still suffer from
substantial artificial helicity drift because small phase errors, divergence
errors, and curl errors accumulate over long time intervals.

To reduce this source of error, we do not learn \(\bomega\) as an independent
network output. Instead, the vorticity used in both the PDE residual and the
helicity diagnostic is computed from the velocity field by automatic
differentiation,
\[
\bomega_\Theta=\nabla\times\bu_\Theta .
\]
This choice enforces the compatibility relation
\[
\nabla\cdot(\nabla\times\bu_\Theta)=0
\]
at the representation level and removes one common source of spurious helicity
production. The same compatibility issue is central in curl-conforming and
geometric discretizations of rotational flow models
\cite{girault2006curl,gawlik2020conservative,hu2021helicity}.

We also monitor the kinetic energy,
\[
E(t) = \int_\Omega \frac12|\bu(t,\bx)|^2\,d\bx .
\]
For the benchmark considered in this work, the prescribed energy level is
\[
E_0=\frac{1}{120}.
\]
Because the reported test is force-free and uses a very large Reynolds number,
the physical viscous energy decay over the simulated interval is negligible
relative to the optimization error of the neural solver. We therefore impose
$E_0$ as an energy-consistency level. For moderate Reynolds numbers or forced
flows, this term should be replaced by the corresponding energy-balance loss.

In the temporal partitioning experiment, the invariant-guided transfer is field
based rather than purely scalar. The terminal velocity snapshot produced on one
time slab is imposed as the initial state for the next slab through a hard
temporal constraint. The trained slab sequence is then used as the teacher for a
residual global space-time correction. The energy level $E_0$ is imposed along
the temporal sequence and during the global correction.

This design is motivated by the structure of helicity itself. Since
\(\cH_f(t)\) depends simultaneously on \(\bu\) and on
\(\nabla\times\bu\), matching the scalar helicity alone is generally
insufficient to control the long-time dynamics. By transferring an
energy-consistent velocity field and computing vorticity from that field, the
method constrains both the magnitude and the rotational structure of the neural
solution. Helicity is therefore monitored as a derived invariant during slab
training and is weakly regularized only in the final residual correction. Its
stability still depends primarily on the curl-compatible velocity
representation, energy-consistent temporal transfer, and global correction.
This is consistent with the broader helicity literature, where the quantity is
viewed as a measure of vortex-line linkage and as a constraint on relaxation
and turbulent transfer
\cite{cantarella1999influence,frisch1975possibility,perez2009role,
arnold2008topological}.

\paragraph{Role in IG-PINN.}
For this problem, IG-PINN uses temporal partitioning to control long-time
error accumulation. The transferable information is not the scalar helicity
alone, but an energy-consistent velocity trace passed between adjacent time
slabs. The vorticity is reconstructed from the transferred velocity field by
automatic differentiation, and the final residual correction produces a single
space-time neural field.

%% file: sections/method.tex
\section{Invariant Guided PINNs}
\label{sec:method}

The model problems in Section~\ref{sec:governing} suggest a common
algorithmic principle. Instead of using partitioned neural fields as the final
piecewise approximation, we use partitioned training stages to generate
physically transferable information and then train a single globally defined
neural field. The transfer may occur in space, as in the Oldroyd--B cylinder
problem, or in time, as in the rotational Newtonian helicity problem. In both
cases, the partitioned stage acts as a conservative preconditioning step for
the final global correction.

We refer to the resulting method as an Invariant Guided PINN
(IG-PINN). The method has three components: local partitioned training,
extraction of transferable physical quantities, and global-field correction.
The specific transfer variables are problem dependent, but the structure of the
algorithm is the same for spatial and temporal partitions.

\begin{figure}[t]
\centering
\includegraphics[width=0.95\linewidth]{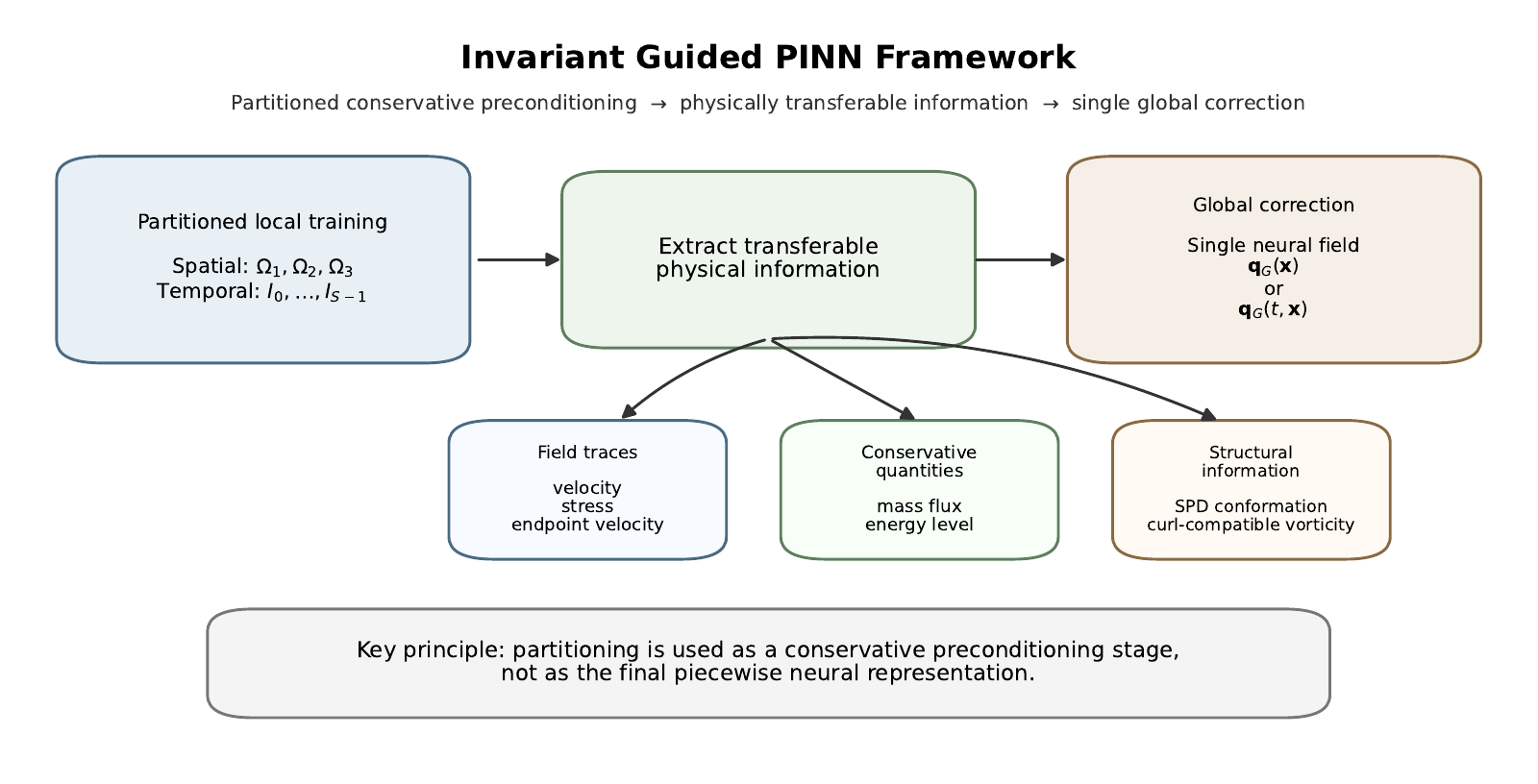}
\caption{Overview of IG-PINN. Partitioned local training is used as a
conservative preconditioning stage. The extracted field traces, conservative
diagnostics, and structural information are then used to train a single global
neural representation.}
\label{fig:ig_framework}
\end{figure}

This distinction separates IG-PINN from standard interface-matching
domain-decomposed PINNs. In cPINN- or XPINN-type formulations, the subdomain
networks usually constitute the final approximation and are coupled through
interface penalties \cite{jagtap2020conservative,jagtap2020extended,
shukla2021parallel,Moseley2023}. In Schwarz-type neural domain decomposition,
subdomain
problems are repeatedly solved while exchanging boundary data until interface
consistency is reached. In contrast, IG-PINN uses the partitioned stage once
to generate physically transferable information. No expensive Schwarz-type
alternating iteration is performed. The final approximation is represented by a
single globally defined neural field refined by the full PDE residual, boundary
conditions, and problem-dependent conservative constraints.

\begin{figure}[t]
\centering
\includegraphics[width=0.95\linewidth]{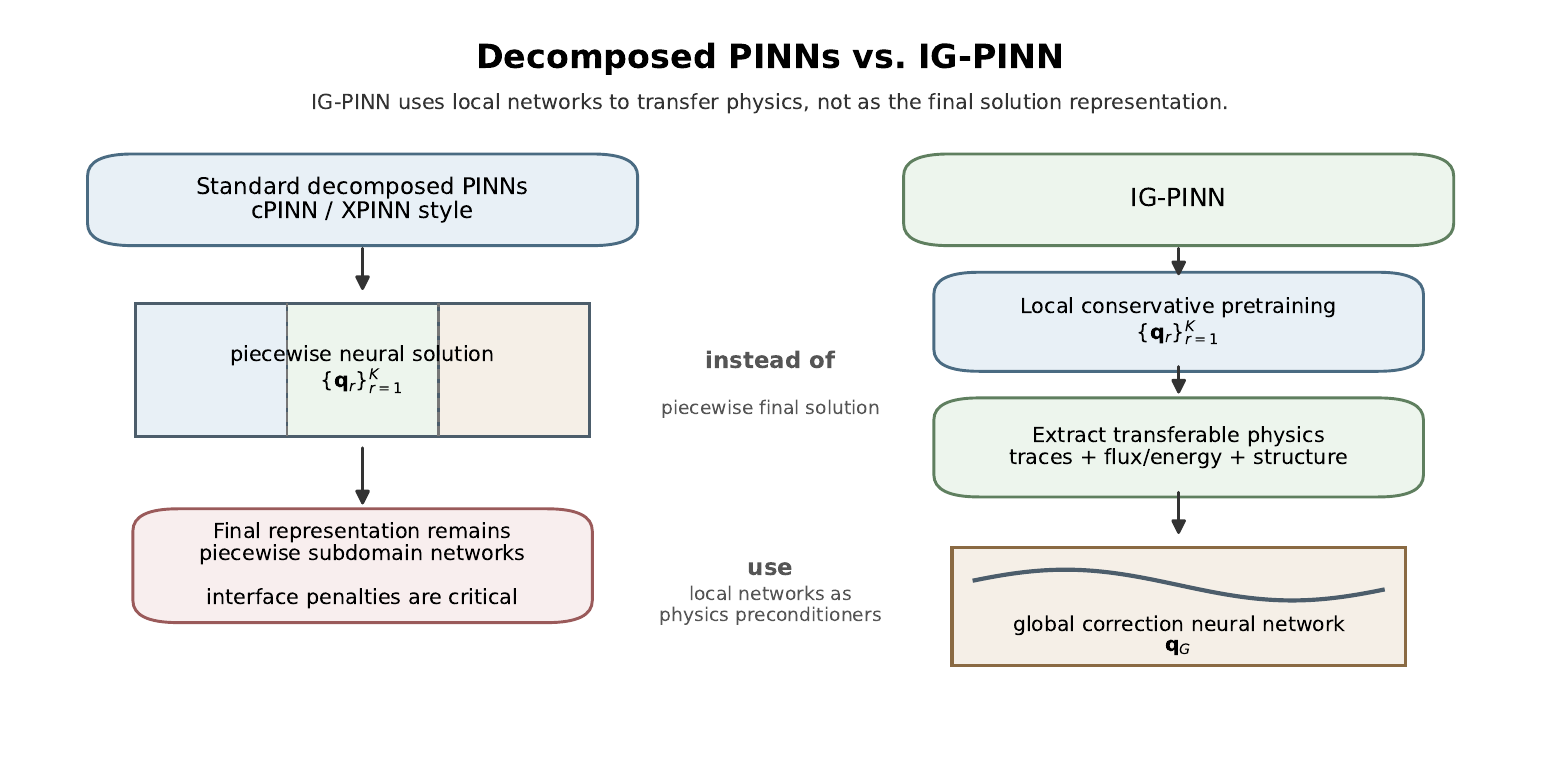}
\caption{Conceptual difference between standard decomposed PINNs and IG-PINN.
The partition-stage neural fields in IG-PINN are not the final solution
representation; they provide transferable physical information for a final
global correction.}
\label{fig:decomposed_vs_ig}
\end{figure}

\subsection{Local partition training}
\label{subsec:partition_training}

Let \(D=\Omega\) for a steady problem and
\(D=[0,T]\times\Omega\) for a transient problem. We cover \(D\) by an ordered
collection of spatial subdomains or temporal slabs,
\[
D=\bigcup_{r=1}^K D_r .
\]
On each \(D_r\), we train a local neural approximation
\[
\bq_r(\bz;\theta_r), \qquad
\bz=\bx
\quad \mbox{or} \quad
\bz=(t,\bx),
\]
for the relevant physical variables. A typical local loss has the form
\begin{equation}
\cL_r(\theta_r) =
\cL_{\mathrm{PDE}}^r
+ \cL_{\mathrm{BC}}^r
+ \cL_{\mathrm{IC}}^r
+ \cL_{\mathrm{tr}}^r
+ \cL_{\mathrm{cons}}^r .
  \label{eq:local_loss}
  \end{equation}
  Some terms are absent depending on the problem. For instance, the steady
  Oldroyd--B problem has no temporal initial condition, whereas the helicity
  problem uses slab-interface information but no stress transfer.

The transfer term \(\cL_{\mathrm{tr}}^r\) is not intended to copy all network
outputs across an interface. In incompressible flow, the pressure is defined
only up to an additive constant unless a gauge is fixed. Passing pressure
values directly between subdomains can therefore introduce a nonphysical
constraint. IG-PINN transfers only variables that carry physically meaningful
interface information, together with integral quantities selected from the
conservation, balance, or structural properties of the governing equations.

\subsection{Trace, conservative, and structural information}
\label{subsec:extraction}

After the local problems have been trained, we extract three classes of
information:
\[
\mathscr{E}_r =
\left(
\{\mathcal{T}_{r,\ell}[\bq_r]\}_{\ell},
\{\cQ_{r,m}[\bq_r]\}_{m},
\{\mathcal{R}_{r,n}[\bq_r]\}_{n}
\right).
\]
Here $\{\mathcal{T}_{r,\ell}\}$ denotes a field trace or sampled solution
operator, $\{\cQ_{r,m}\}$ denotes an integral diagnostic or conservative scalar,
and $\{\mathcal{R}_{r,n}\}$ denotes structural information such as admissibility
or compatibility constraints.

In the Oldroyd--B problem, the transferred traces are
\[
\mathcal{T}_{r,\ell}[\bq_r] =
(u_r,v_r,\tau_{xx,r},\tau_{xy,r},\tau_{yy,r})
\quad \mbox{on selected interfaces}.
\]
The pressure is excluded from the transfer, and a pressure gauge is imposed
only in the global correction. Since \(\btau\) is an affine function of the
conformation tensor \(\bA\), transferring \(\btau\) is equivalent to
transferring the non-isotropic conformation information needed by the stress
balance. The mass fluxes
\[
\cM_r(x_j) =
\int_{\Gamma_{x_j}} u_r(x_j,y)\,dy
\]
are used as conservative diagnostics. The positive-definiteness of the
conformation tensor is treated as a structural admissibility constraint rather
than as a conserved quantity.

In the helicity problem, the temporal transfer is field based. The primary
trace is the terminal velocity of each slab,
\[
\mathcal{T}_s[\bq_s]=\bu_s(t_{s+1},\cdot).
\]
The conservative scalar imposed along the temporal sequence is the prescribed
energy level $E_0$. The helicity $\cH_f(t)$ is monitored a posteriori, but
it is not imposed as an independent scalar transfer constraint. Instead, it is
stabilized indirectly through the transferred velocity field, the
curl-compatible definition
\[
\bomega_\Theta=\nabla\times\bu_\Theta,
\]
and the energy-consistency constraint.

\subsection{Global-field correction}
\label{subsec:global_correction}

The extracted traces and conservative quantities are used to build a
partition-informed base field and then to train a globally defined residual
correction. This residual form is the variant used in the long-time helicity
experiments. Let \(\bq_B\) denote the field supplied by the partition stage. It
may be a teacher-guided full-domain approximation constructed from the spatial
subdomain predictions, or a frozen time-slab teacher operator evaluated on the
appropriate temporal slab. Instead of replacing this field by a freely
initialized global PINN, we write the corrected field as
\begin{equation}
\bq_G(\bz;\Theta_R)
=
\bq_B(\bz)
+ \gamma \chi(\bz)\boldsymbol{N}_R(\bz;\Theta_R),
\qquad
\bz=\bx
\quad \mbox{or} \quad
\bz=(t,\bx).
\label{eq:residual_global_correction}
\end{equation}
Here \(\boldsymbol{N}_R\) is the trainable residual network, \(\gamma>0\) is a
residual scale, and \(\chi\) is an optional gate used to preserve hard
constraints that are already satisfied by \(\bq_B\). For the helicity problem,
\(\bq_B=\bq_S\) is the hard-time slab teacher and
\(\chi(t,\bx)=t/T\), so the initial velocity remains fixed at \(t=0\). For a
steady spatial problem, \(\chi\) may be chosen as one or as a boundary-compatible
factor when hard boundary values should be left unchanged. In implementation,
the last layer of \(\boldsymbol{N}_R\) can be initialized to zero, so the global
model starts exactly from \(\bq_B\) and learns only the correction required by
the full-domain equations.

The residual corrector may be newly initialized, or it may reuse compatible
weights from the partition stage. The essential point is that the final trainable
object is the residual correction in \eqref{eq:residual_global_correction},
while the partition output serves as a fixed base trajectory or base field.
The global correction is still implemented as a two-stage procedure. Stage I
uses teacher replay together with the physical and conservative losses to keep
the correction near the transferred partition solution while removing obvious
gaps or accumulated defects. Stage II drops the teacher or interface-data term
and continues training the same residual model using only the physical
residuals, boundary conditions, conservative constraints, and structural
constraints.

\begin{figure}[H]
\centering
\includegraphics[width=0.95\linewidth]{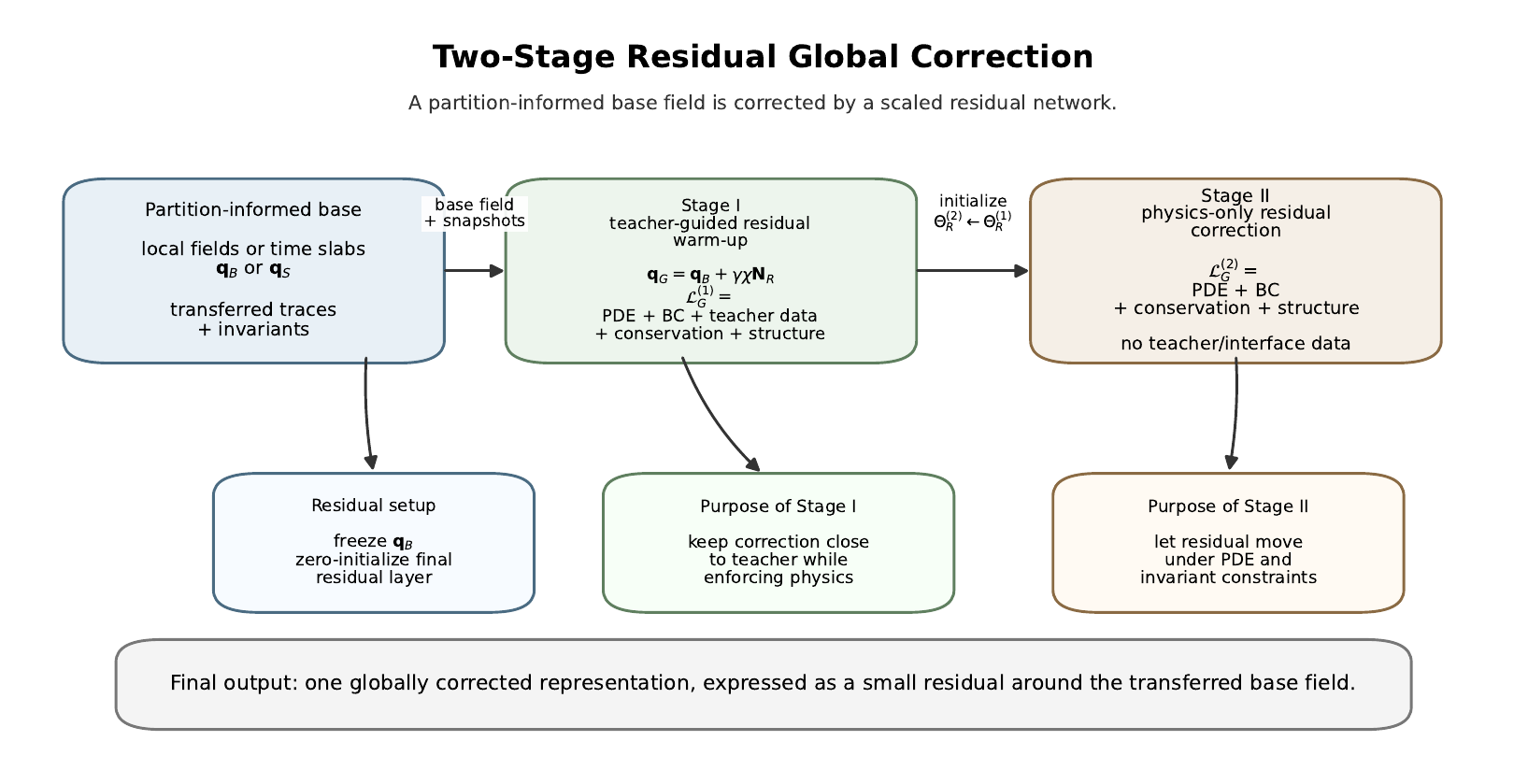}
\caption{Two-stage residual global correction. A partition-informed base field
is corrected by a scaled residual network. Stage I uses partition-teacher data,
conservative quantities, and physics rules as a teacher-guided warm-up. Stage
II removes the teacher or interface-data term and continues the residual
correction using only the PDE, boundary, conservative, and structural
constraints.}
\label{fig:two_stage_global_correction}
\end{figure}

Let \(\cD_T\) denote the teacher data extracted from the partition stage. For
the Oldroyd--B problem, \(\cD_T\) consists of sampled velocity and stress
predictions from the streamwise subdomain models. For the helicity problem,
\(\cD_T\) consists of sampled velocity, vorticity, and divergence snapshots
from the trained time-slab sequence. With the residual parameterization
\eqref{eq:residual_global_correction}, the Stage-I correction is obtained by
minimizing
\begin{equation}
\begin{aligned}
\cL_G^{(1)}(\Theta_R)
&=
\cL_{\mathrm{PDE}}^G
+ \cL_{\mathrm{BC}}^G
+ \lambda_T \cL_T^G
+ \lambda_{\mathrm{cons}}
  \cL_{\mathrm{cons}}^G
+ \lambda_{\mathrm{str}}
  \cL_{\mathrm{str}}^G .
  \end{aligned}
  \label{eq:global_stage1_loss}
  \end{equation}
Here
\[
\cL_T^G =
\frac{1}{|\cD_T|}
\sum_{(\bz_j,\boldsymbol{y}_j^T)\in \cD_T}
\left|
\mathcal{P}[\bq_G](\bz_j)-\boldsymbol{y}_j^T
\right|^2
\]
is the teacher-data loss. The observation operator \(\mathcal{P}\) selects the
quantities that are physically meaningful to replay: velocity and stress for
the Oldroyd--B transfer, and velocity, vorticity, and divergence for the
helicity residual correction. The conservative loss has the generic form
\[
\cL_{\mathrm{cons}}^G =
\sum_m
\left|
\cQ_m[\bq_G]-\overline{\cQ}_m
\right|^2 .
\]
  The reference value \(\overline{\cQ}_m\) is chosen according to the
  available physical information. For the Oldroyd--B cylinder problem, the
  reference mass flux is the prescribed inlet flux $\cM_0$. For the helicity
  benchmark, the reference energy is the prescribed level $E_0$. The structural
  term $\cL_{\mathrm{str}}^G$ represents problem-dependent constraints such as
  SPD conformation parameterization or curl-compatible vorticity reconstruction.

After Stage I, the residual parameters are used as initialization for the final
global correction,
\[
\Theta_R^{(2,0)}=\Theta_R^{(1)} .
\]
The teacher term is then dropped:
\begin{equation}
\cL_G^{(2)}(\Theta_R)
=
\cL_{\mathrm{PDE}}^G
+ \cL_{\mathrm{BC}}^G
+ \lambda_{\mathrm{cons}}\cL_{\mathrm{cons}}^G
+ \lambda_{\mathrm{str}}\cL_{\mathrm{str}}^G .
\label{eq:global_stage2_loss}
\end{equation}
Thus the final global correction does not train a second unrelated global
solution. Stage I learns a small residual around the partition-informed base
field while replaying physically meaningful teacher quantities; Stage II lets
that residual move under the full PDE, boundary conditions, and conservative or
structural constraints, without continuing to copy the partition teachers.

The role of partitioning in IG-PINN is therefore different from that in
standard decomposed PINNs. In cPINN and XPINN, partitioning defines the
representation of the final solution. In IG-PINN, partitioning defines a
conservative preconditioning stage for the optimization problem. The local
networks are discarded after their physically transferable information has
been extracted.

\subsection{Spatial realization for the Oldroyd--B problem}
\label{subsec:oldroyd_realization}

For the non-Newtonian cylinder problem, we use a streamwise partition defined
by two artificial interfaces \(\xi_1<\xi_2\):
\begin{align*}
\Omega_1 &= \Omega_{\mathrm{OB}}\cap\{x<\xi_1\},\\
\Omega_2 &= \Omega_{\mathrm{OB}}\cap\{\xi_1<x<\xi_2\},\\
\Omega_3 &= \Omega_{\mathrm{OB}}\cap\{x>\xi_2\}.
\end{align*}

\begin{figure}[t]
\centering
\includegraphics[width=0.95\linewidth]{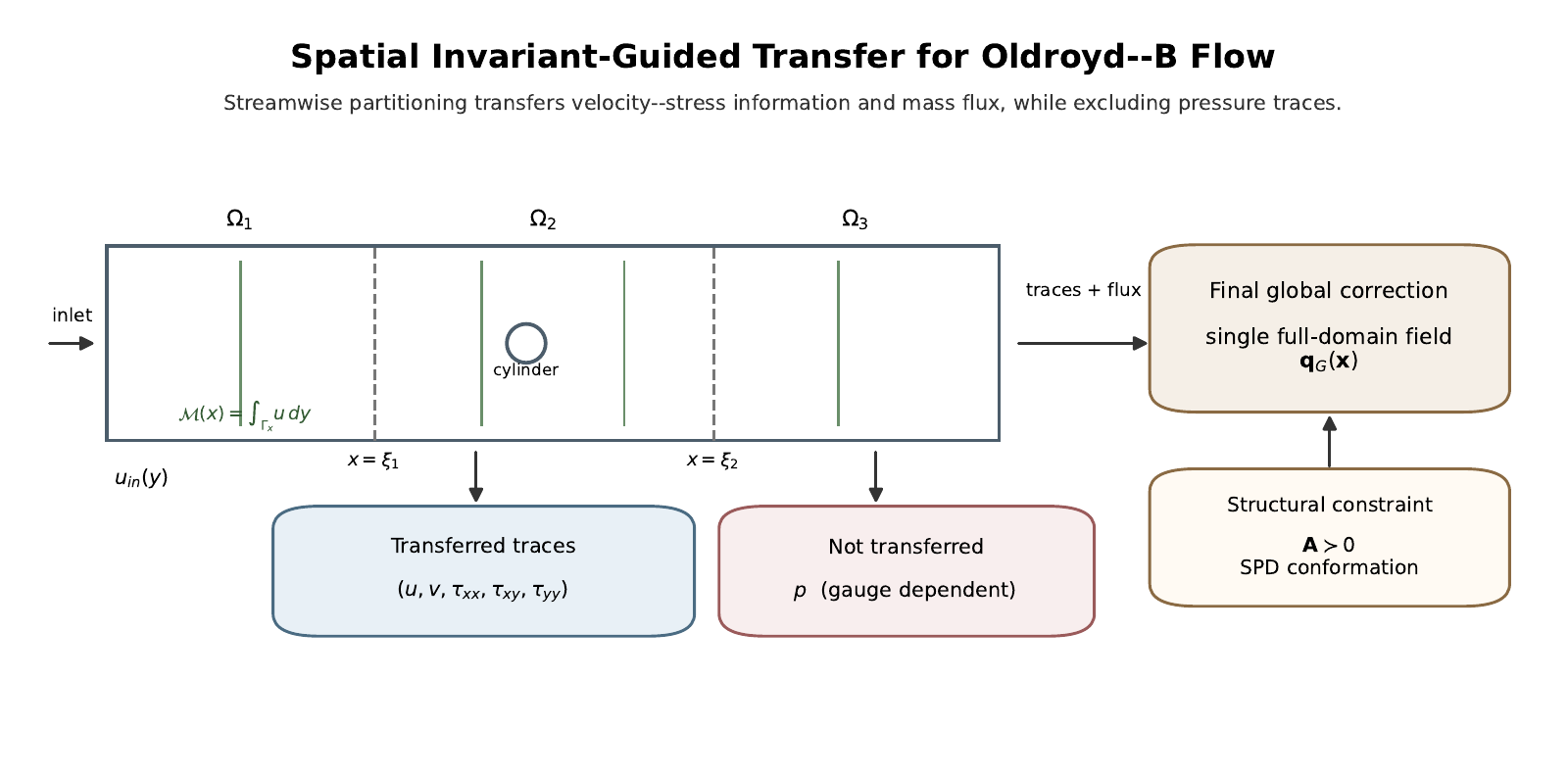}
\caption{Spatial invariant-guided transfer for the Oldroyd--B cylinder problem.
The local stage transfers velocity--stress traces and mass-flux information,
while pressure traces are excluded because of gauge non-uniqueness.}
\label{fig:oldroyd_spatial_transfer}
\end{figure}

The shared network is trained sequentially from inlet to outlet. At the
artificial interfaces \(x=\xi_1\) and \(x=\xi_2\), the receiving stage uses the
upstream-stage prediction of
\[
(u,v,\tau_{xx},\tau_{xy},\tau_{yy})
\]
as physically transferable interface data. The pressure is not transferred
across artificial interfaces.

The conformation tensor is represented through an SPD-preserving
parameterization, so that \(\bA\) remains positive definite during training.
This structural constraint is important for the Oldroyd--B model because loss
of positive definiteness may lead to nonphysical polymeric stresses and
unstable residuals, a central issue in high-Weissenberg-number viscoelastic
simulation \cite{baaijens1998mixed,fattal2005time,lozinski2003energy,
balci2011symmetric}.

After all subdomain stages are trained, their interior predictions are sampled
and used to construct the base field \(\bq_B\) for the full-domain correction.
In practice this base field can be obtained by a teacher-guided full-domain
approximation to the transferred velocity--stress predictions while also
enforcing the Oldroyd--B residuals, inlet condition, no-slip boundary condition,
SPD conformation parameterization, pressure gauge, and flux loss
\eqref{eq:flux_loss}. The final correction then uses the residual form
\eqref{eq:residual_global_correction}: the residual network starts from a small
or zero correction to \(\bq_B\), and the teacher-data term is removed in the
second stage so that the remaining update is driven by the physical residuals,
boundary conditions, flux conservation, pressure gauge, and SPD structure. The
final output is a single neural field over the whole fluid domain rather than a
piecewise collection of subdomain networks.

\subsection{Temporal realization for the helicity problem}
\label{subsec:helicity_realization}

For the transient helicity problem, the time interval is partitioned into slabs
\[
I_s=[t_s,t_{s+1}],
\qquad
s=0,\ldots,S-1 .
\]
On each slab, the network output is
\[
\bq_s(t,\bx)=(\bu_s(t,\bx),\widetilde{p}_s(t,\bx)),
\]
while
\[
\bomega_s(t,\bx)=\nabla\times\bu_s(t,\bx)
\]
is obtained by automatic differentiation. The slab is local only in time. Each
slab model is global in space, and no spatial partition-of-unity assembly is
used in the helicity experiment.

The left temporal interface is imposed by a hard output transform. Let
\[
\eta_s(t)=\frac{t-t_s}{t_{s+1}-t_s},
\qquad
0\leq \eta_s(t)\leq 1,
\]
and let the raw network output be
\[
(\boldsymbol{N}_s^u,N_s^p).
\]
We set
\begin{equation}
\bu_s(t,\bx) =
\boldsymbol{g}_s(t,\bx)
+
\eta_s(t)\boldsymbol{N}_s^u(t,\bx),
\qquad
\widetilde{p}_s(t,\bx)=N_s^p(t,\bx),
\label{eq:helicity_hard_time}
\end{equation}
where
\[
\boldsymbol{g}_0(t,\bx)=\bu_0(\bx),
\]
and, for \(s\geq 1\),
\[
\boldsymbol{g}_s(t,\bx) =
\operatorname{sg}
\left[
\bu_{s-1}(t,\bx)
\right].
\]
Here \(\operatorname{sg}[\cdot]\) denotes a stop-gradient operation. Since
\(\eta_s(t_s)=0\), the interface velocity is imposed exactly and no separate
initial-condition penalty is used in the hard-time runs. This follows the same
principle as hard-constraint neural constructions for boundary and interface
conditions \cite{lu2021hard,sukumar2022exact}.

The energy-consistency constraint is imposed at selected previous slab
endpoints and at selected points in the current slab:
\begin{equation}
\cL_E^{(s)} =
\frac{1}{|\mathcal{S}_s|}
\sum_{\tau\in\mathcal{S}_s}
\left(
\frac{E_s(\tau)-E_0}{E_0+\varepsilon}
\right)^2 .
\label{eq:helicity_energy_loss}
\end{equation}
The complete slab loss used in the conservative run is
\begin{equation}
\cL_{\mathrm{slab}}^{(s)} =
\cL_{\mathrm{PDE}}^{(s)}
+ w_{\mathrm{bc}}\cL_{\mathrm{BC}}^{(s)}
+ w_E\cL_E^{(s)} .
  \label{eq:helicity_slab_loss}
  \end{equation}

\begin{figure}[t]
\centering
\includegraphics[width=0.95\linewidth]{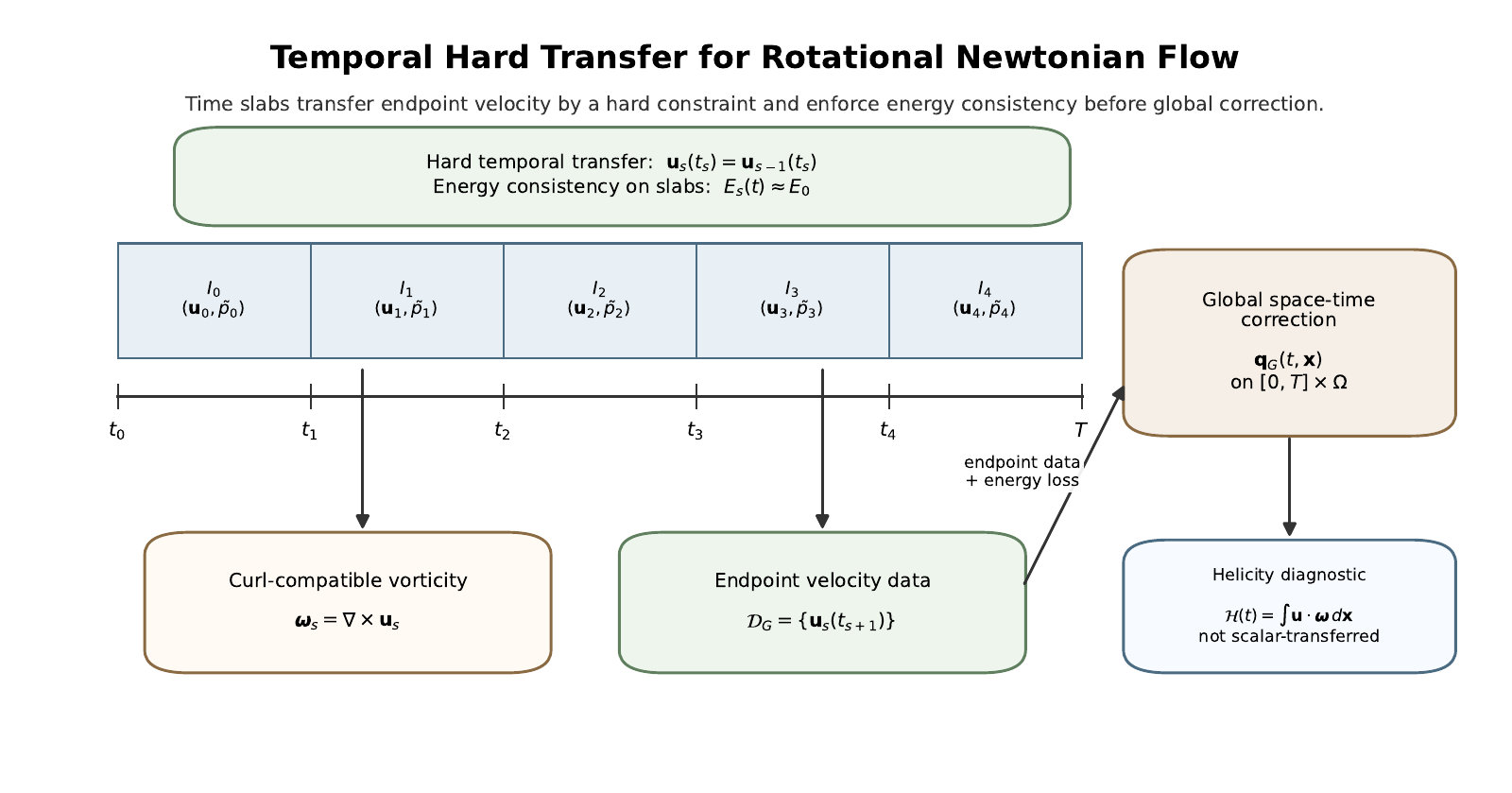}
\caption{Temporal invariant-guided transfer for the rotational Newtonian flow. The
left endpoint of each slab is imposed by a hard velocity transform, energy is
controlled along the time sequence, and the trained slab sequence is used as a
teacher for the residual global space-time correction.}
\label{fig:helicity_temporal_transfer}
\end{figure}

After all slabs are trained, we perform a final global correction analogous
to the full-domain correction in the Oldroyd--B experiment. In the helicity
experiment, the effective variant is a residual correction around the trained
hard-time slab sequence. Let \(\bq_S(t,\bx)\) denote the piecewise time-slab
teacher obtained from the trained slab models. The global model is
\[
\bq_G(t,\bx;\Theta_G) =
\bq_S(t,\bx)+\gamma\frac{t}{T}\boldsymbol{N}_G(t,\bx;\Theta_G),
\qquad
\bq_G=(\bu_G,\widetilde{p}_G),
\]
where \(\gamma>0\) is a residual scale. The factor \(t/T\) keeps the initial
velocity fixed because the slab teacher already satisfies the hard initial
condition at \(t=0\). The residual output layer is initialized to zero in the
reported implementation, so the global correction starts exactly from the
piecewise hard-time slab teacher and learns only the space-time residual needed
to improve the structure diagnostics. Thus the final representation is a single
space-time network correction on \([0,T]\times\Omega\), but it starts from the
conservative slab trajectory rather than from a randomly initialized global
field.

Let \(\cD_G\) denote sampled slab-teacher snapshots over the whole time interval.
Each snapshot contains velocity, vorticity, and divergence values,
\[
(\tau,\bx_j,\bu_j^S,\bomega_j^S,d_j^S),
\qquad
\bomega_j^S=\nabla\times\bu_S(\tau,\bx_j),
\qquad
d_j^S=\nabla\cdot\bu_S(\tau,\bx_j).
\]
The global data loss uses these field-level quantities:
\begin{align}
\cL_{\mathrm{data}}^G
&=
\frac{1}{|\cD_G|}
\sum_{(\tau,\bx_j)\in\cD_G}
\left[
\frac{\left|\bu_G-\bu_j^S\right|^2}{2E_0+\varepsilon}
+ w_\omega
\frac{\left|\nabla\times\bu_G-\bomega_j^S\right|^2}{2E_0+\varepsilon}
+ w_d
\frac{\left|\nabla\cdot\bu_G-d_j^S\right|^2}{\sigma_d^2}
\right],
\label{eq:helicity_global_data_loss}\\
\cL_E^G
&=
\frac{1}{|\mathcal{S}_G|}
\sum_{\tau\in\mathcal{S}_G}
\left(
\frac{E_G(\tau)-E_0}{E_0+\varepsilon}
\right)^2,
\label{eq:helicity_global_energy_loss}\\
\cL_H^G
&=
\frac{1}{|\mathcal{S}_H|}
\sum_{\tau\in\mathcal{S}_H}
\left(
\frac{\cH_f^G(\tau)-\cH_f(0)}{\sigma_H}
\right)^2,
\label{eq:helicity_global_helicity_loss}\\
\cL_{G,\mathrm{time}}^{(1)}
&=
\cL_{\mathrm{PDE}}^G
+ w_{\mathrm{bc}}\cL_{\mathrm{BC}}^G
+ w_{\mathrm{data}}\cL_{\mathrm{data}}^G
+ w_E\cL_E^G
+ w_H\cL_H^G .
  \label{eq:helicity_global_loss}
  \end{align}
  Here \(\sigma_d\) and \(\sigma_H\) are normalization scales for the divergence
  replay and helicity regularization terms.
  In the slab stage, helicity is monitored as a diagnostic rather than used as
  the transferred scalar. In the final residual correction, the weak term
  \(\cL_H^G\) stabilizes the derived helicity while the field-level data loss
  is used only during the Stage-I warm-up to keep the residual correction close
  to the curl-compatible slab trajectory. Stage II initializes from the Stage-I
  residual correction and removes \(\cL_{\mathrm{data}}^G\):
  \begin{equation}
  \cL_{G,\mathrm{time}}^{(2)}
  =
  \cL_{\mathrm{PDE}}^G
  + w_{\mathrm{bc}}\cL_{\mathrm{BC}}^G
  + w_E\cL_E^G
  + w_H\cL_H^G .
  \label{eq:helicity_global_stage2_loss}
  \end{equation}
  This second stage keeps the same residual global parameterization but no
  longer forces the space-time model to copy the slab teacher.

\subsection{Algorithmic summary}
\label{subsec:algorithm}

The constructions above can be summarized as an invariant-guided
optimization strategy. The partitioned stages are not treated as the final
piecewise numerical solution. They are used to reduce the stiffness of the
original large-domain or long-time training problem and to expose physical
information that can be transferred safely. The transferred information is deliberately
selective: it includes traces that are well defined across the partition,
integral conservative quantities that encode the relevant balance law, and
structural constraints that keep the neural field in the physically admissible
state space.

For the Oldroyd--B cylinder wake, this means transferring velocity and stress
traces together with mass-flux information, while excluding pressure traces
because the pressure gauge is not physically transferable. These data define a
partition-informed base field; the final residual correction then drops the
teacher-data term and enforces the Oldroyd--B residuals, no-slip and inlet
constraints, the flux balance, the pressure gauge, and the SPD conformation
parameterization. For the helicity problem, the same logic is applied in time:
each slab transfers its endpoint velocity through a hard-time constraint, the
energy level is imposed as the conservative scalar, helicity is monitored during
slab training, and the two-stage residual global correction produces one
differentiable representation on \([0,T]\times\Omega\).

Thus IG-PINN differs from cPINN and XPINN in the role assigned to
decomposition. Standard decomposed PINNs use the subdomain networks as the
solution representation and rely on interface matching to assemble a piecewise
field. IG-PINN uses decomposition as a conservative preconditioning stage and
returns a single global neural field after correction. Algorithm~\ref{alg:ig}
summarizes this workflow.

\begin{algorithm}[H]
\caption{Invariant Guided PINN}
\label{alg:ig}
\begin{algorithmic}[1]
\State Choose a spatial partition, temporal partition, or space-time partition \(\{D_r\}_{r=1}^K\).
\For{\(r=1,\ldots,K\)}
\State Train or continue a partition-stage PINN \(\bq_r(\cdot;\theta_r)\) on \(D_r\) using \eqref{eq:local_loss}.
\State If \(D_r\) receives information from an earlier partition, impose only physically transferable traces.
\EndFor
\For{\(r=1,\ldots,K\)}
\State Extract field traces \(\mathcal{T}_{r,\ell}[\bq_r]\), conservative diagnostics \(\cQ_{r,m}[\bq_r]\), and structural information \(\mathcal{R}_{r,n}[\bq_r]\).
\EndFor
\State Build a partition-informed base field \(\bq_B\) from the extracted teacher information.
\State Define \(\bq_G=\bq_B+\gamma\chi\boldsymbol{N}_R(\cdot;\Theta_R)\) on the full spatial domain or full space-time.
\State Stage I: train the residual corrector using teacher data, PDE residuals, boundary conditions, conservative constraints, and structural constraints.
\State Stage II: initialize from Stage I, remove the teacher-data term, and continue training the residual corrector using only physical, conservative, and structural constraints.
\State \Return the globally corrected IG-PINN solution.
\end{algorithmic}
\end{algorithm}

\clearpage

%% file: sections/numerical_experiments.tex
\section{Numerical Experiments}
\label{sec:numerics}

Section~\ref{sec:method} described IG-PINN as an invariant-guided
procedure. The purpose of the numerical
experiments is to test the two realizations of that procedure rather than to
compare two identical neural architectures. The Oldroyd--B cylinder wake tests
spatial invariant-guided transfer: local streamwise subdomains are trained first,
velocity--stress traces and mass-flux information are extracted, and a
full-domain correction produces the reported flow field. The helicity problem
tests temporal invariant-guided transfer: time slabs pass endpoint velocity through
a hard-time construction, the energy level is used as the conservative scalar,
and a final space-time correction gives one global representation.

This organization mirrors Algorithm~\ref{alg:ig}. Both experiments therefore
share the same algorithmic protocol: partitioned local training, selective
transfer of physically meaningful information, final global correction, and
diagnostics for residual error, conservative drift, structural admissibility,
and field quality. The network sizes are problem dependent because the two
systems have different state variables and dimensions. Table~\ref{tab:numerical_protocol}
summarizes the common protocol and the problem-specific choices.

\begin{table}[H]
\centering
\footnotesize
\setlength{\tabcolsep}{3pt}
\renewcommand{\arraystretch}{0.92}
\begin{tabular}{>{\raggedright\arraybackslash}p{0.20\linewidth}
                >{\raggedright\arraybackslash}p{0.36\linewidth}
                >{\raggedright\arraybackslash}p{0.36\linewidth}}
\toprule
Experiment & Partition and transferred information & Final global correction and diagnostics \\
\midrule
Oldroyd--B cylinder wake &
Sequential training on three streamwise subdomains; one shared 12-by-80
two-dimensional PINN transfers velocity--stress traces, mass flux, and SPD
admissibility without transferring pressure &
Full-domain global correction with transferred predictions, PDE and boundary
constraints, flux conservation, and SPD structure; the second stage removes
teacher data and continues with physical, conservative, and structural
constraints; diagnostics report drag, residual, flux error, wall speed, and
SPD violations \\
Hard-time helicity transfer &
Twenty temporal slabs with $\Delta t=0.5$ on one global spatial domain; slab and
correction networks use six 512-wide hidden layers; transfer uses hard endpoint
velocity, energy target $E_0=1/120$, and slab snapshots &
Full space-time global correction with slab-teacher data, rotational residuals,
boundary constraints, energy consistency, and weak helicity regularization; the
second stage removes slab-teacher data and continues with the residual,
boundary, energy, and helicity terms; diagnostics report energy, helicity,
divergence, and relative field difference
\\
\bottomrule
\end{tabular}
\caption{Unified numerical protocol. The two neural architectures differ
because the Oldroyd--B problem is a steady two-dimensional velocity, pressure,
and stress system, whereas the helicity problem is a transient
three-dimensional velocity and pressure system. Both experiments follow the same
invariant-guided transfer workflow.}
\label{tab:numerical_protocol}
\end{table}

All experiments are implemented in PyTorch and trained with automatic
differentiation on CUDA. The reported runs were carried out on a Linux
workstation with two AMD EPYC 7302 CPUs, 64 GB system memory, and one NVIDIA
Quadro RTX 5000 GPU with 16 GB memory. The software environment used Python
3.12.3, PyTorch 2.11.0, and CUDA 13.0.

\subsection{Non-Newtonian Cylinder Wake}
\label{subsec:oldroyd_results}

The computational domain for the reported Oldroyd--B run is
\[
\Omega_{\mathrm{OB}}
= [-15,15]\times[-2,2]\setminus \overline{B_1(0)} ,
\]
with a unit cylinder centered at the origin. The streamwise IG partition is
\begin{align*}
\Omega_1 &= [-15,-2]\times[-2,2],\\
\Omega_2 &= [-2,5]\times[-2,2]\setminus \overline{B_1(0)},\\
\Omega_3 &= [5,15]\times[-2,2],
\end{align*}
so the artificial transfer interfaces are \(x=-2\) and \(x=5\). At the inlet
\(x=-15\), the prescribed velocity is
\[
u_{\mathrm{in}}(y)=1.5\left(1-\frac{y^2}{4}\right),
\qquad v_{\mathrm{in}}(y)=0,
\]
which gives the reference mass flux \(\cM_0=4\). The experiment uses
\(Re=0\), \(\beta_s=0.59\), and \(Wi=0.1\). The inlet and outlet subdomains use
3000 residual points each, the cylinder subdomain and final full-domain
correction use 10000 residual points, and the full-domain network has 12 hidden
layers with width 80.

\begin{figure}[H]
\centering
\begin{subfigure}{0.85\linewidth}
  \centering
  \includegraphics[width=\linewidth]{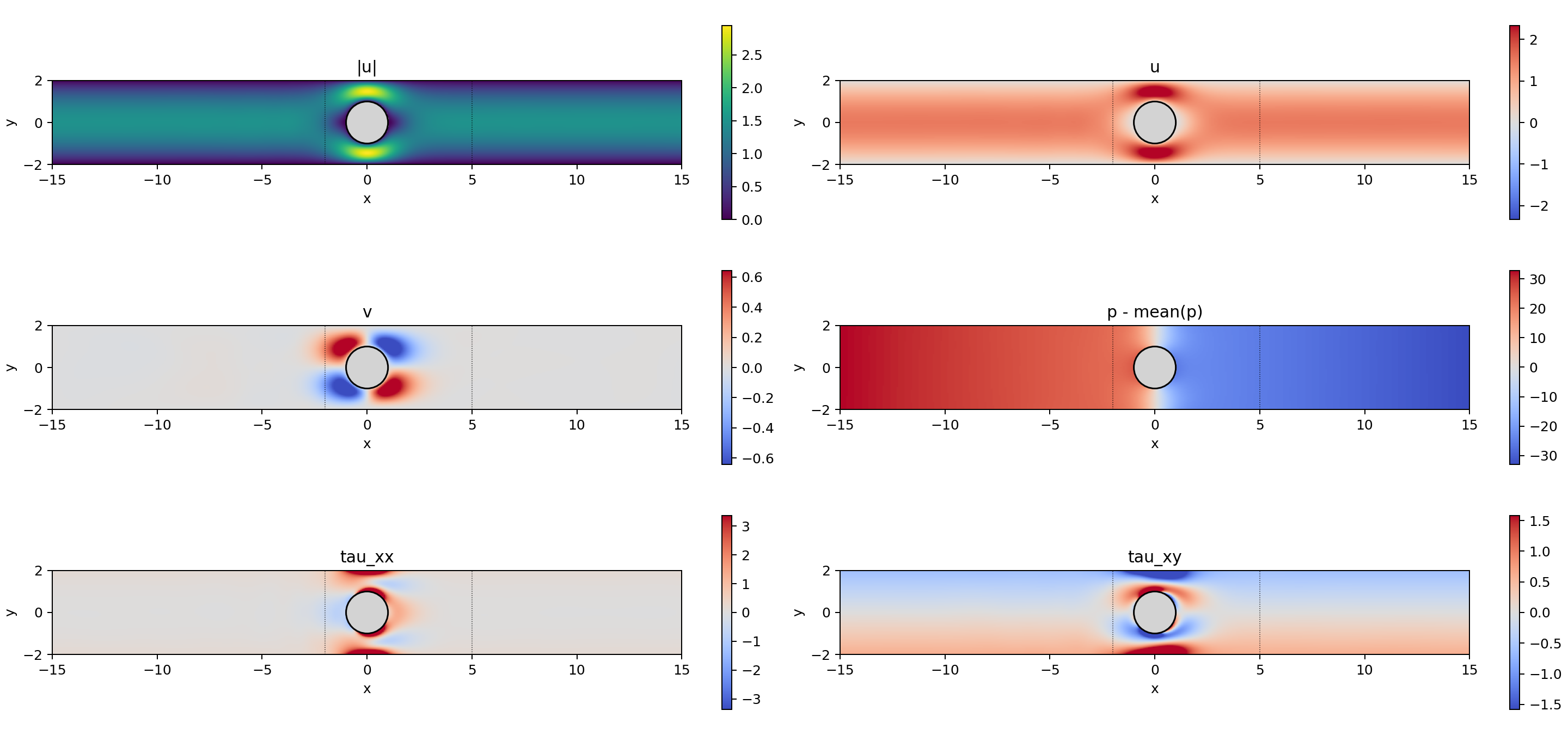}
  \caption{Predicted velocity, pressure, and stress fields.}
\end{subfigure}

\begin{subfigure}{0.85\linewidth}
  \centering
  \includegraphics[width=\linewidth]{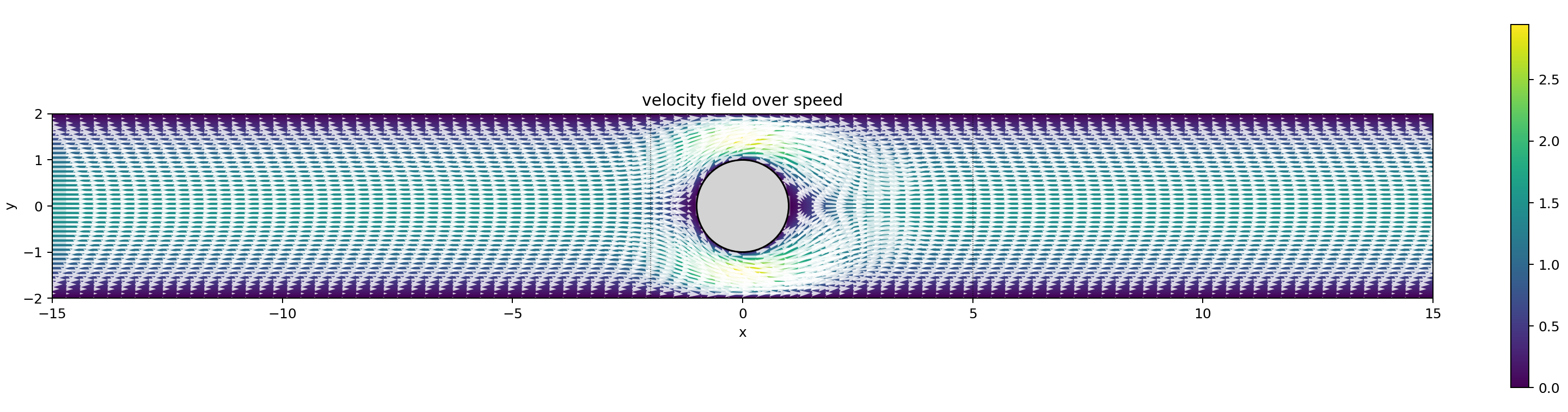}
  \caption{Velocity field over speed magnitude with quiver arrows.}
\end{subfigure}
\caption{Oldroyd--B cylinder wake at $Wi=0.1$ computed by the
\IG{}-PINN workflow.}
\label{fig:oldroyd_fields}
\end{figure}

\begin{figure}[H]
\centering
\begin{subfigure}{0.85\linewidth}
  \centering
  \includegraphics[width=\linewidth]{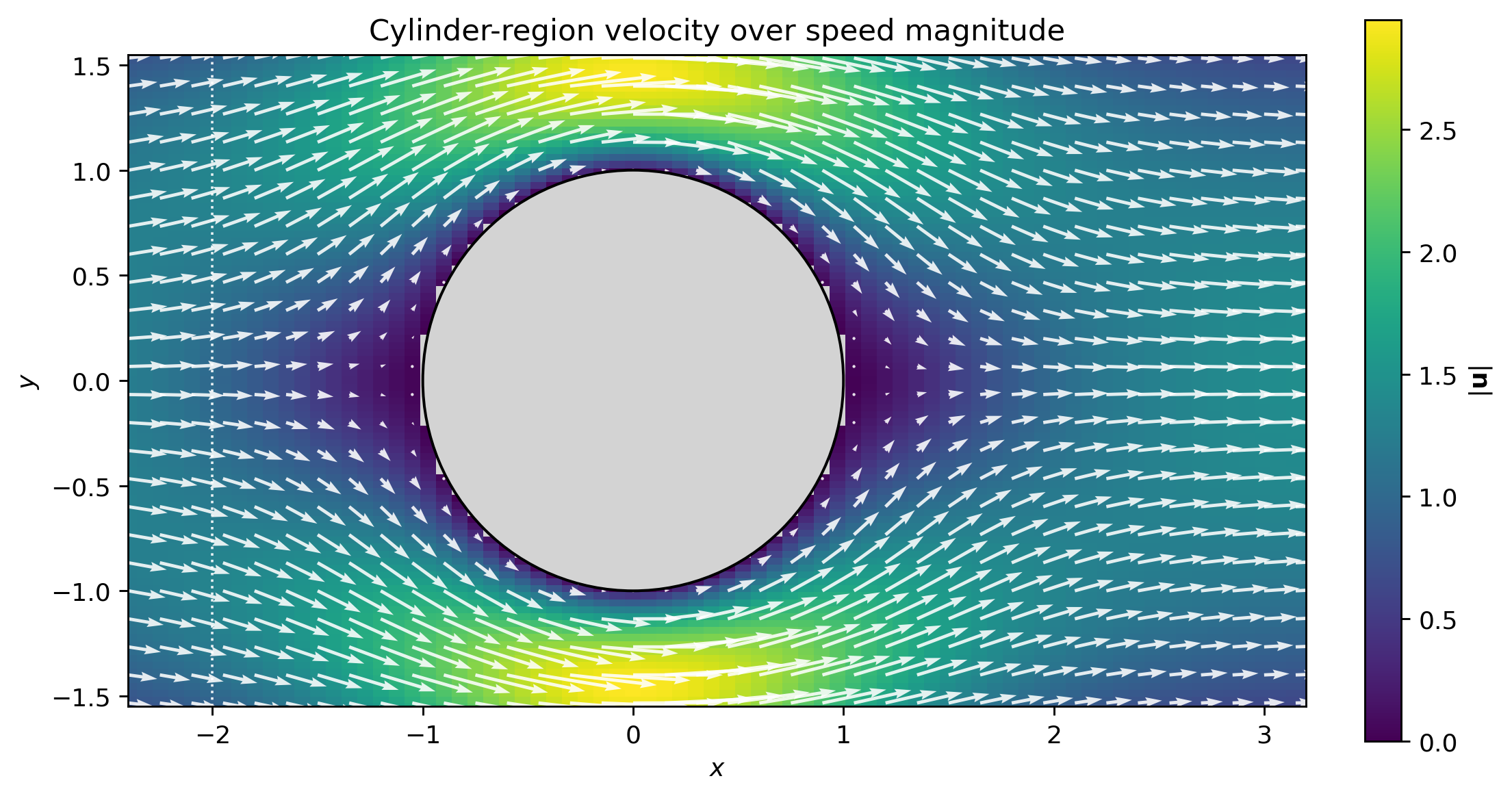}
  \caption{Cylinder-region speed magnitude and velocity arrows.}
\end{subfigure}

\begin{subfigure}{0.85\linewidth}
  \centering
  \includegraphics[width=\linewidth]{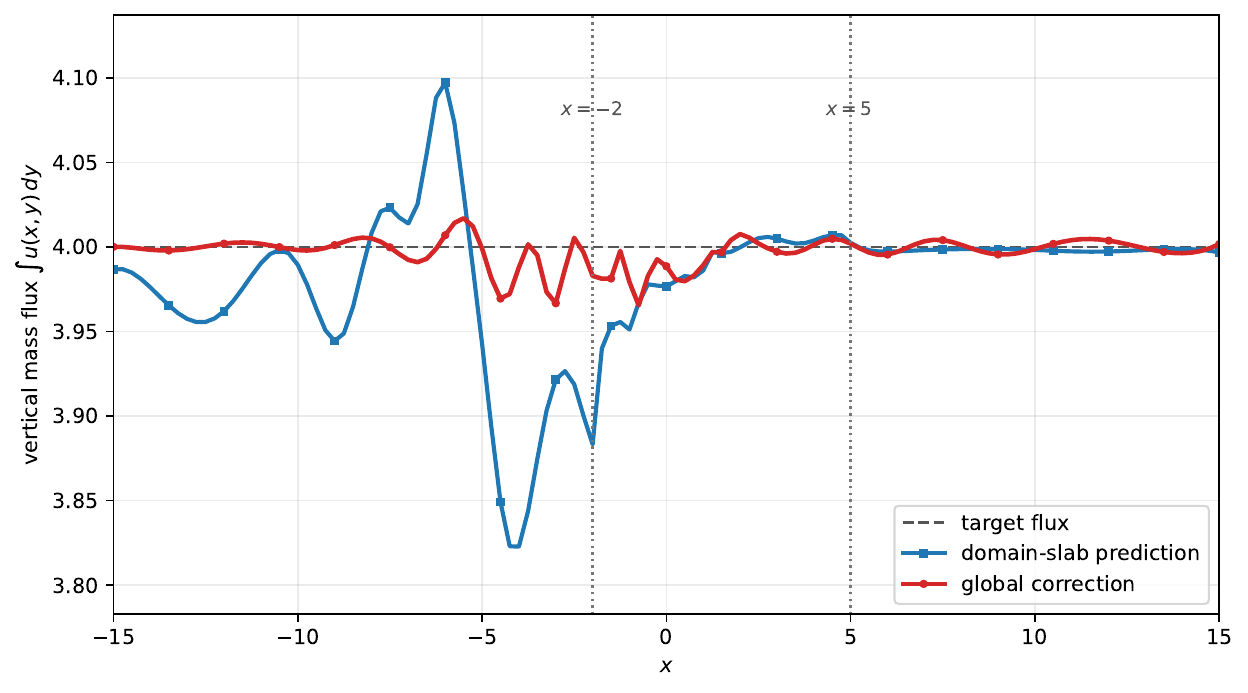}
  \caption{Vertical mass-flux conservation over \(x\in[-15,15]\): stitched
  domain-slab prediction versus global correction.}
\end{subfigure}
\caption{Cylinder-region velocity field and flux-conservation comparison for
the Oldroyd--B cylinder wake.}
\label{fig:oldroyd_zoom_diagnostics}
\end{figure}

Figure~\ref{fig:oldroyd_fields} shows the full-domain prediction for the
Wi=0.1 cylinder wake. The velocity and stress fields remain smooth across the
two subdomain transfer interfaces, and the quiver plot confirms that the
streamwise acceleration and wake recovery are captured by the final global
field. Mass-flux conservation is examined separately in
Figure~\ref{fig:oldroyd_zoom_diagnostics}, where the stitched domain-slab
prediction is compared with the global correction.

Figure~\ref{fig:oldroyd_zoom_diagnostics} magnifies the cylinder region and
compares mass-flux conservation before and after the global correction. The
zoomed velocity plot shows that the near-cylinder acceleration and the
wake-side velocity recovery are resolved more clearly than in the full-channel
quiver plot. The flux comparison is evaluated on vertical sections over
\(x\in[-15,15]\). The domain-slab curve stitches the left, cylinder, and right
local-stage predictions, whereas the global-correction curve is produced by a
single full-domain model. The global correction removes much of the
interface-induced flux variation and keeps the vertical mass flux closer to
the target value \(\cM_0=4\). Since no external exact Oldroyd--B solution is
available for this confined-cylinder run, the reported field diagnostics are
residual and conservation diagnostics rather than an exact-solution error.

Table~\ref{tab:oldroyd_diagnostics} reports the detailed diagnostics for the
final Oldroyd--B global correction model at Wi=0.1. The mean flux is
$3.9969$, close to the target $4$, and the hard boundary operator drives
the wall velocity to machine precision. The minimum sampled eigenvalue of the
conformation tensor stays positive, so the SPD constraint is preserved.

\begin{table}[H]
\centering
\begin{tabular}{lc}
\toprule
Diagnostic & Value \\
\midrule
Drag coefficient $C_D$ & $1.3002\times10^{2}$ \\
Random PDE residual & $3.1469\times10^{-3}$ \\
Mean flux & $3.9969$ \\
Mean absolute flux error & $6.1003\times10^{-3}$ \\
Maximum absolute flux error & $3.4530\times10^{-2}$ \\
Maximum cylinder wall speed & $8.8585\times10^{-7}$ \\
Minimum conformation eigenvalue & $5.1568\times10^{-1}$ \\
SPD violations & $0$ \\
\bottomrule
\end{tabular}
\caption{Diagnostics for the final Wi=0.1 Oldroyd--B global correction model.}
\label{tab:oldroyd_diagnostics}
\end{table}

Table~\ref{tab:oldroyd_comparison} compares the current \IG{} run with the
baseline PINN, XPINN, and cPINN observations from the earlier non-Newtonian
comparison study. These baseline entries are used only to quantify the failure
modes of direct PINN or direct decomposed PINN training. A vanilla PINN
converges to a near-trivial interior velocity field. XPINN does not recover the
cylinder wake reliably, and cPINN gives a more structured field but still
exhibits interface discontinuities and a drag value around $74$. By contrast,
the current \IG{} run uses sequential invariant-guided transfer followed by a
final full-domain global correction. The benchmark drag is not treated
as a single exact value: Table~1 of Goyal and Derksen \cite{goyal2012direct}
reports, for the closely related De$=0.1$ Oldroyd--B cylinder benchmark,
values ranging from the Alves et al. value $129.91$ to $131.73$ on their
finest grid, with coarser-grid values extending higher. The present run uses
$Wi=0.1$, $\beta_s=0.59$, and $Re=0$, whereas the tabulated case uses
De$=0.1$, $\beta_s=0.6$, and $Re=0.067$, so the table is used as a benchmark
band rather than as a unique ground truth.

\begin{table}[H]
\centering
\resizebox{\linewidth}{!}{%
\begin{tabular}{>{\raggedright\arraybackslash}p{0.19\linewidth}
                >{\raggedright\arraybackslash}p{0.28\linewidth}
                >{\raggedright\arraybackslash}p{0.37\linewidth}c}
\toprule
Method & Conservative or transfer mechanism & Observed behavior at $Wi=0.1$ & $C_D$ \\
\midrule
Vanilla PINN & PDE residual and boundary terms only &
The predicted flow collapses to a near-zero interior state away from the
inlet. & -- \\
XPINN & Solution and residual continuity across subdomain interfaces &
The wake is not recovered reliably and visible interface artifacts remain. & -- \\
cPINN & Conservative interface coupling &
The field is better than XPINN but discontinuities persist; the reported drag is
far below the reference. & $\approx 74$ \\
\IG{}-PINN Stage-I transfer-guided global approximation & Sequential
trace/stress/flux transfer and teacher-guided full-domain approximation &
Smooth global field; random PDE residual $8.9343\times10^{-3}$ and mean flux
error $9.7347\times10^{-3}$. & $130.534$ \\
\IG{}-PINN Stage-II final global correction & Full-domain correction under PDE,
boundary, flux, and SPD constraints & Random PDE residual $3.1469\times10^{-3}$,
mean flux error $6.1003\times10^{-3}$, and no sampled SPD violation. & $130.022$ \\
Alves et al. \cite{alves2001flow} & Literature benchmark &
Reference value & $129.91$ \\
\bottomrule
\end{tabular}
}
\caption{Oldroyd--B confined-cylinder comparison at $Wi=0.1$. The PINN,
XPINN, and cPINN entries summarize the earlier baseline comparison. The
\IG{} rows are evaluated from the current code run. The literature values are
used as a benchmark interval because the tabulated De$=0.1$ case has slightly
different $Re$ and viscosity ratio.}
\label{tab:oldroyd_comparison}
\end{table}

The comparison shows that the improvement is not obtained merely by adding
interfaces. The important step is to transfer conservative information that is
physically meaningful for the next stage and then remove the remaining
piecewise artifacts through a global correction. The final global correction
reduces the independent random PDE residual by about $65\%$ and the mean flux
error by about $37\%$ relative to the Stage-I transfer-guided global
approximation, while keeping the drag inside the Alves-to-M3
benchmark interval in Table~\ref{tab:oldroyd_comparison}. Relative to the
Alves et al. value $129.91$ reported in Goyal and Derksen's table, the final
\IG{} drag differs by about $0.09\%$.

\subsection{Hard-Time Energy-Constrained Helicity Transfer}
\label{subsec:helicity_transfer_results}

For the Newtonian helicity experiment we use the rotational formulation
\eqref{eq:lamb_momentum}--\eqref{eq:lamb_div}. The force-free hard-time test
uses the spatial domain \(\Omega_{\mathrm{H}}=[0,1]^3\), Reynolds number
$Re=10^{12}$, total time $T=10$, slab size $\Delta t=0.5$, and a final
global space-time correction over the full interval. The initial velocity is
\[
\begin{aligned}
u_1(0,x,y,z) &= -\sin(\pi(x-0.5))\cos(\pi(y-0.5))z(z-1),\\
u_2(0,x,y,z) &= \phantom{-}\cos(\pi(x-0.5))\sin(\pi(y-0.5))z(z-1),\\
u_3(0,x,y,z) &= 0.
\end{aligned}
\]
The helicity experiment follows the same conservative narrative as the
Oldroyd--B wake, but the partition direction is time. First, each temporal slab
is trained as a local-in-time problem with one global spatial PINN. Second, the
velocity output is written in the hard-time form \eqref{eq:helicity_hard_time},
so the left endpoint of each slab exactly inherits the previous slab state.
Third, the slab stage enforces the energy target $E_0=1/120$ through
\eqref{eq:helicity_energy_loss}. Fourth, the piecewise hard-time sequence is
used as a teacher for a residual-form global correction
\[
\bq_G(t,\bx)=\bq_S(t,\bx)+0.05\,\frac{t}{T}\boldsymbol{N}_G(t,\bx).
\]
This final stage is analogous to the full-domain Oldroyd--B correction: the
global model remains close to the transferred local solution, while the
rotational PDE, boundary conditions, energy consistency, and weak helicity
regularization reduce accumulated long-time defects.

Table~\ref{tab:helicity_settings} lists the numerical protocol for the
hard-time run. The temporal interface is imposed by construction. Dense teacher
snapshots are extracted from the slab sequence over the full time interval;
these snapshots contain velocity, vorticity, and divergence values and are used
only to keep the global correction tied to the conservative slab trajectory.

\begin{table}[H]
\centering
\resizebox{\linewidth}{!}{%
\begin{tabular}{lc}
\toprule
Setting & Value \\
\midrule
Time interval & $0\leq t\leq10$, $S=20$, $\Delta t=0.5$ \\
Network per slab and global correction & $(4,512,512,512,512,512,512,4)$ \\
Hard temporal interface & Endpoint transfer by \eqref{eq:helicity_hard_time} \\
Conservative quantity & Kinetic energy target \(E_0=1/120\) \\
Stage-I teacher data & Velocity, vorticity, and divergence snapshots from the slab sequence \\
Stage-II correction & Teacher data removed; rotational residuals, boundary conditions, energy consistency, and weak helicity regularization retained \\
\bottomrule
\end{tabular}
}
\caption{Numerical setup for the force-free helicity transfer experiment. The
conservative workflow performs a final global correction after all time slabs
have been trained.}
\label{tab:helicity_settings}
\end{table}

\begin{figure}[H]
\centering
\includegraphics[width=0.95\linewidth]{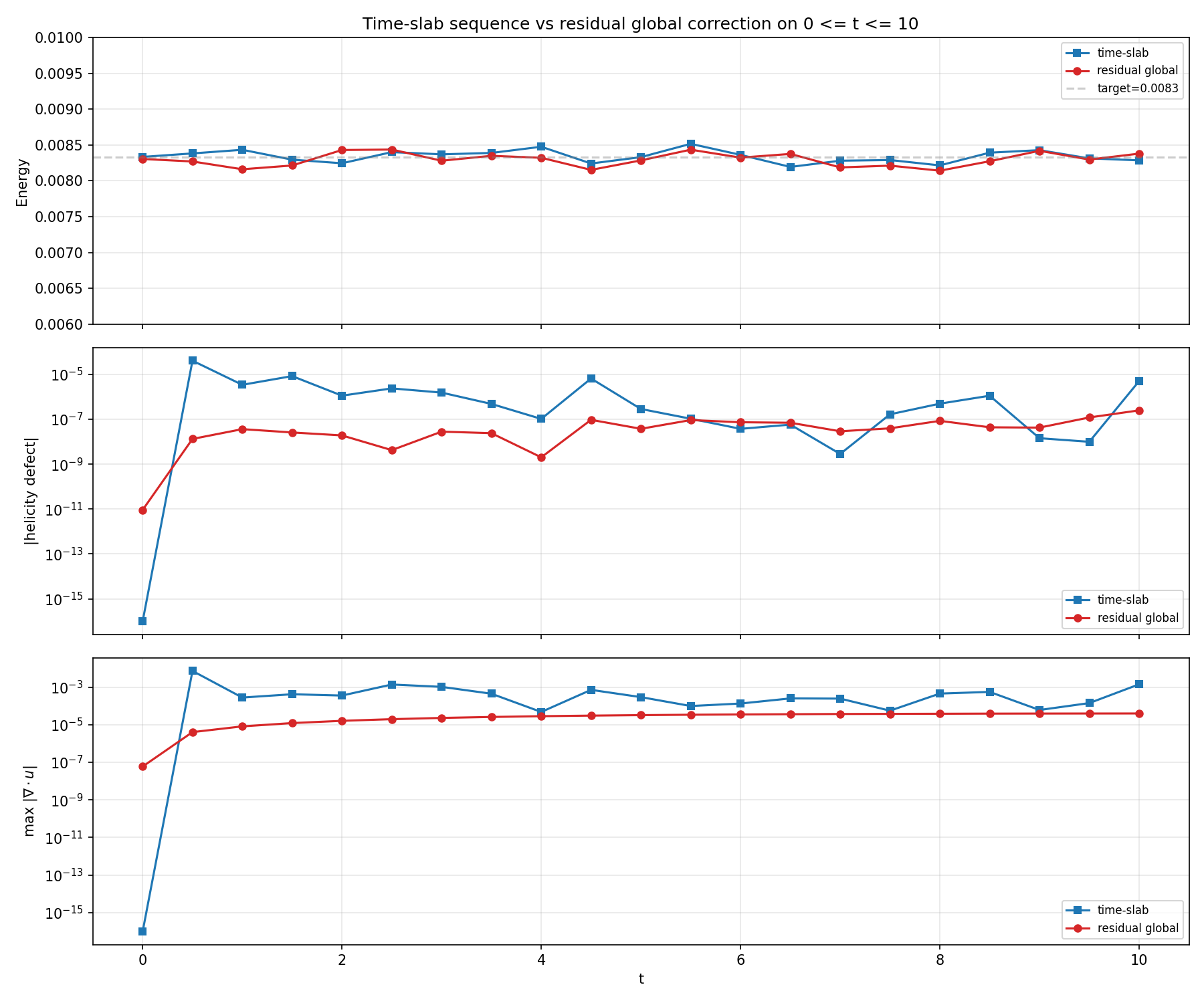}
\caption{Time-slab sequence versus final global correction for the
force-free helicity benchmark over $0\leq t\leq10$. The final global correction
keeps the final field within $0.54\%$ relative \(L^2\) of the slab teacher while
substantially reducing the maximum helicity and divergence defects, showing that
the final global correction is effective at preserving the invariant and
divergence structure of the flow.}
\label{fig:helicity_transfer_defects}
\end{figure}

Figure~\ref{fig:helicity_transfer_defects} compares the original hard-time slab
sequence with the final global correction over the full interval. The energy
panel reports the sampled kinetic energy against the target \(E_0=1/120\). The
helicity and divergence panels report the absolute helicity defect and maximum
sampled incompressibility defect. The last panel reports the relative \(L^2\)
difference between the global correction model and the time-slab teacher at the
sampled endpoint times. Thus the figure measures both conservation improvement
and how far the final single space-time representation moves away from the slab
trajectory. The comparison shows that the global correction is especially
effective as a structure-preserving stage: it retains the transported velocity
field while reducing the accumulated helicity and incompressibility defects of
the slab sequence.

Table~\ref{tab:helicity_transfer} reports the sampled conservation diagnostics.
At the final time \(T=10\), the slab sequence has an energy defect of
$-4.8466\times10^{-5}$ and a helicity defect of $5.0462\times10^{-6}$. The
final global correction changes the final energy defect to
$4.5606\times10^{-5}$ and reduces the final helicity defect to
$2.4985\times10^{-7}$. Over all sampled endpoint times, the maximum helicity
defect decreases from $4.0745\times10^{-5}$ to $2.4985\times10^{-7}$, and the
maximum divergence defect decreases from $7.2350\times10^{-3}$ to
$3.9540\times10^{-5}$. The maximum relative \(L^2\) difference from the slab
teacher is only $5.3931\times10^{-3}$, so the correction improves the
structural diagnostics without replacing the transported velocity trajectory by
an unrelated field.

\begin{table}[H]
\centering
\resizebox{\linewidth}{!}{%
\begin{tabular}{lcc}
\toprule
Diagnostic & Slab sequence & Final global correction \\
\midrule
Time interval & $0\leq t\leq10$ & $0\leq t\leq10$ \\
Final energy defect & $-4.8466\times10^{-5}$ & $4.5606\times10^{-5}$ \\
Maximum absolute energy defect & $1.8230\times10^{-4}$ & $1.9256\times10^{-4}$ \\
Final helicity defect & $5.0462\times10^{-6}$ & $2.4985\times10^{-7}$ \\
Maximum absolute helicity defect & $4.0745\times10^{-5}$ & $2.4985\times10^{-7}$ \\
Maximum divergence defect & $7.2350\times10^{-3}$ & $3.9540\times10^{-5}$ \\
Maximum relative \(L^2\) difference to slab teacher & -- & $5.3931\times10^{-3}$ \\
\bottomrule
\end{tabular}
}
\caption{Force-free hard-time helicity diagnostics. The slab sequence preserves
the nontrivial energy level through conservative energy replay; the final
global correction converts the time-slab sequence into a single space-time PINN
while reducing helicity and divergence drift.}
\label{tab:helicity_transfer}
\end{table}

\begin{figure}[H]
\centering
\begin{subfigure}{0.85\linewidth}
  \centering
  \includegraphics[width=\linewidth]{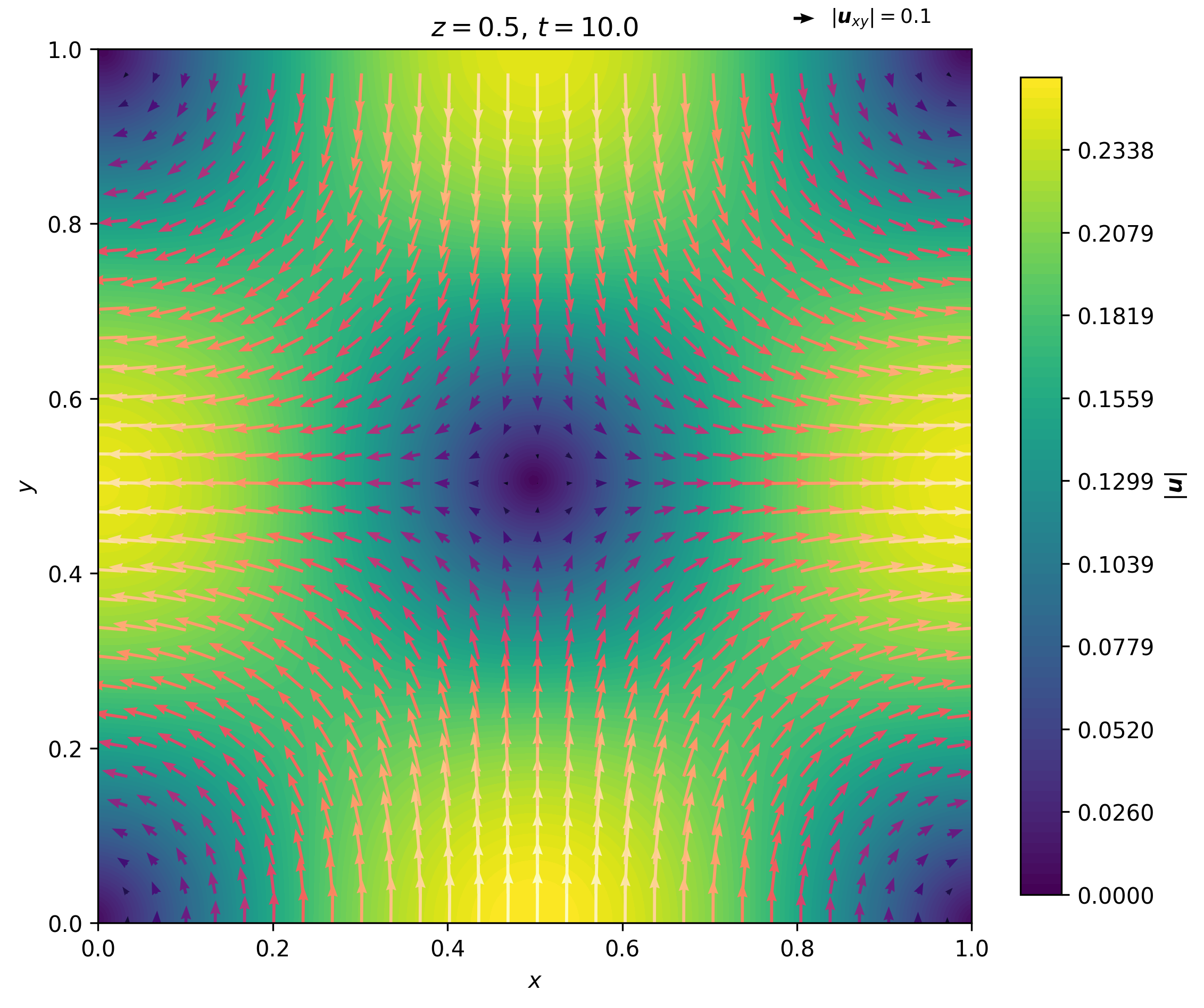}
  \caption{Mid-plane velocity arrows at \(z=0.5\), coloured by speed magnitude.}
\end{subfigure}

\begin{subfigure}{0.48\linewidth}
  \centering
  \includegraphics[width=\linewidth]{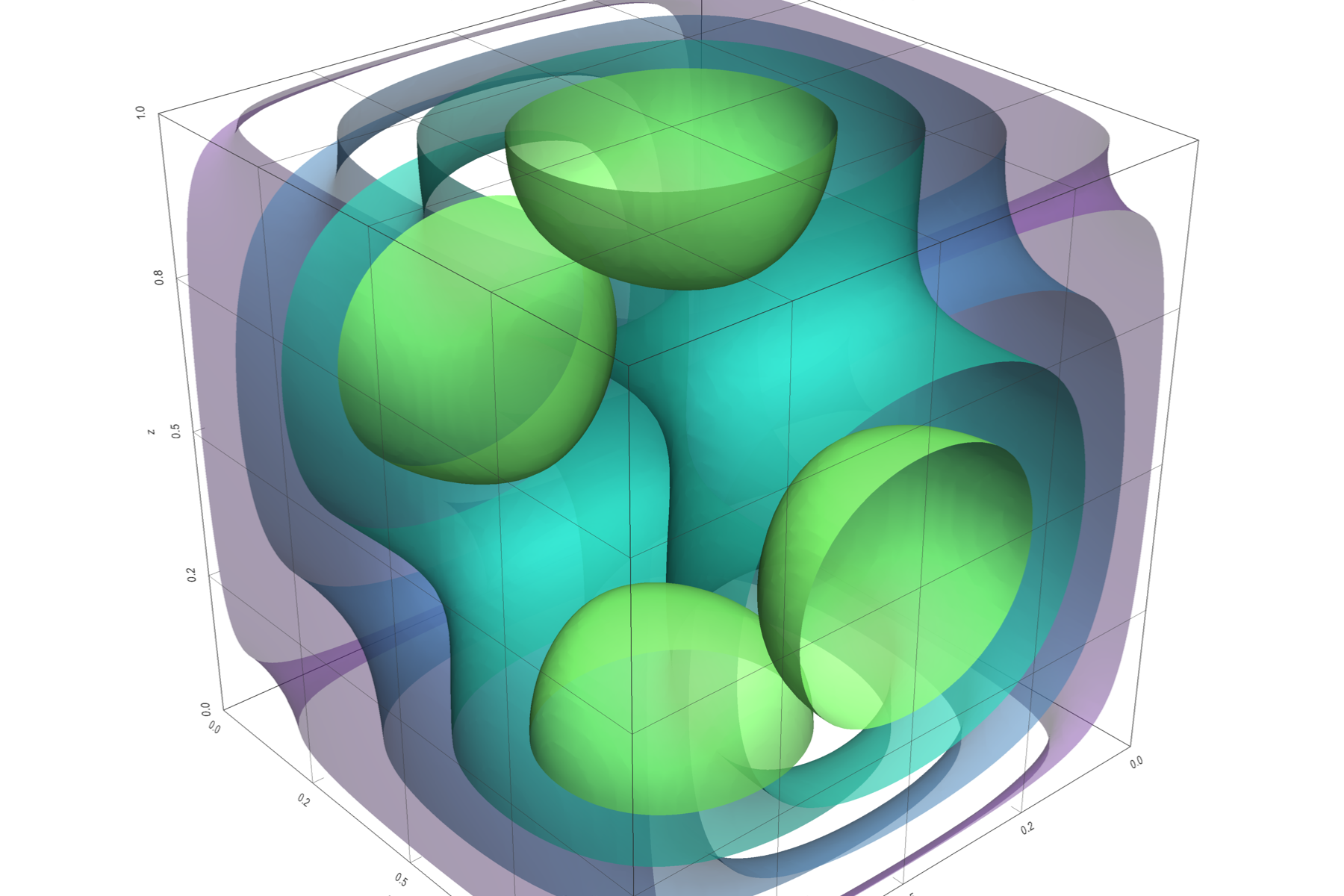}
  \caption{Speed iso-surfaces.}
\end{subfigure}
\hfill
\begin{subfigure}{0.48\linewidth}
  \centering
  \includegraphics[width=\linewidth]{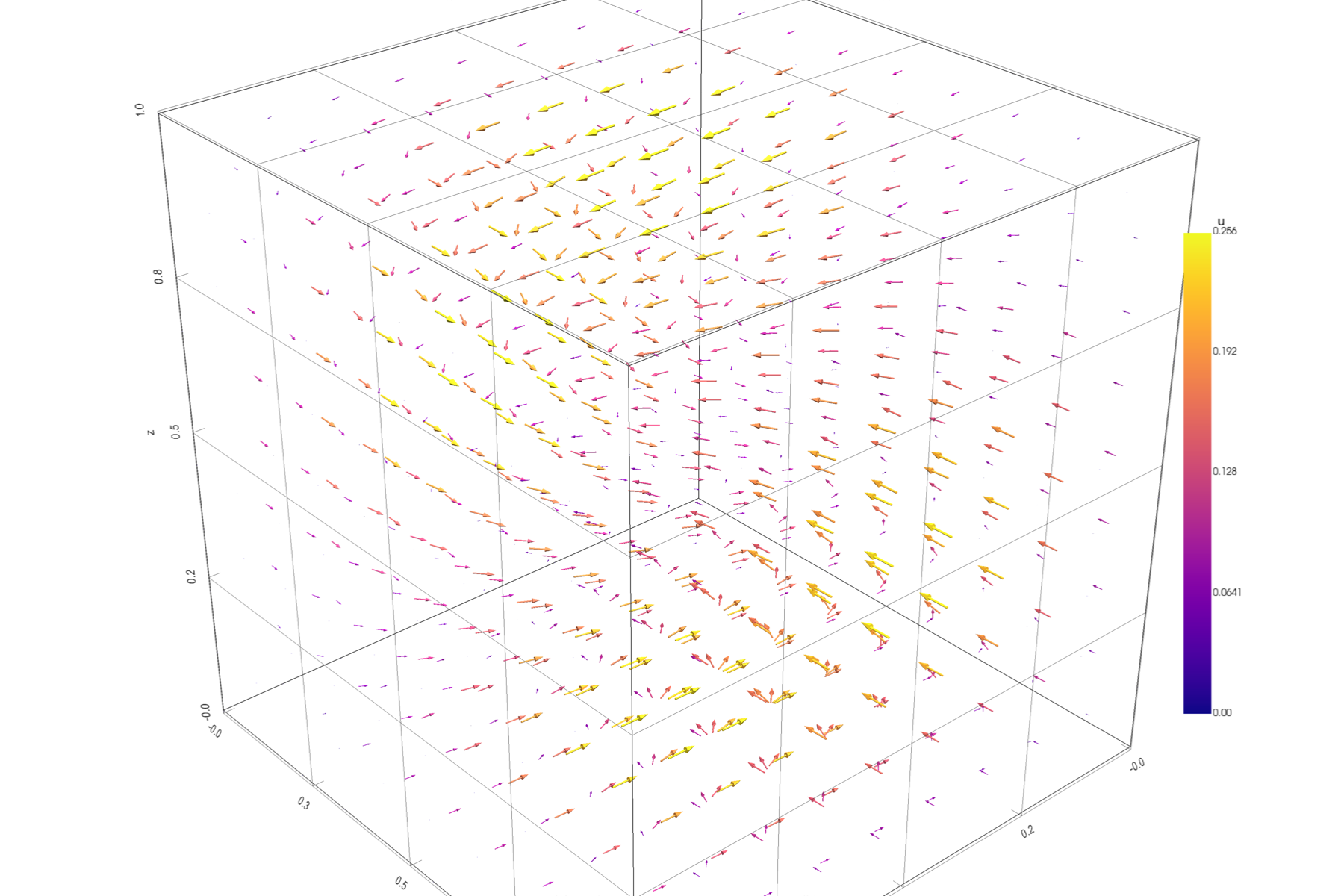}
  \caption{Sparse velocity arrows coloured by speed.}
\end{subfigure}
\caption{Velocity-field views of the global-correction helicity solution at
\(t=10\).}
\label{fig:helicity_terminal_3d}
\end{figure}

Figure~\ref{fig:helicity_terminal_3d} gives complementary views of the globally
corrected velocity field at the reported final time \(t=10\). The \(z=0.5\)
mid-plane plot shows the in-plane velocity direction over the speed magnitude.
The iso-surfaces display the nested high-speed regions, while the sparse
three-dimensional arrows show that the vortex orientation remains coherent
throughout the spatial domain.

This experiment shows why the transferred quantity must be a field-level state
rather than only a scalar invariant. The IG transfer propagates a
curl-compatible velocity field, controls its kinetic energy, and computes
helicity from the induced vorticity. The final global correction then
converts the sequence of conservative slab solutions into one space-time
representation while keeping the correction small in relative \(L^2\), mirroring
the final full-domain correction used in the non-Newtonian cylinder experiment.

%% file: sections/discussion.tex
\section{Discussion}
\label{sec:discussion}

The two examples use different conservative quantities, but the same algorithmic
principle. In the Oldroyd--B cylinder wake, the physically transferable
information is the velocity-stress trace and the mass flux. Pressure is
deliberately excluded from the interface transfer because its additive constant
is not physically determined. The final full-domain correction then removes
subdomain artifacts while retaining flux conservation and SPD conformation
structure. In the helicity problem, the decomposition is temporal rather than
spatial: each slab uses a global spatial PINN, and the transferable information
is a velocity field snapshot together with the energy level. After the slab
sequence has been trained, the complete time-slab trajectory acts as a teacher
for a residual global space-time correction. Computing
$\bomega=\nabla\times\bu$ ensures that the helicity diagnostic is tied to the
same velocity field being propagated across time slabs and then corrected
globally.

This perspective also clarifies the difference between domain decomposition and
invariant-guided transfer. Decomposition reduces the size of
the optimization problem, but the global physical solution emerges only after
local information is converted into the correct conservative variables and
reintroduced into a global correction. The proposed framework can therefore be
viewed as a conservative data-assimilation layer placed between local PINN
training and final global PINN correction. This layer has the same role in both
experiments: it turns local predictions into physically meaningful data for the
last whole-domain optimization.

Several limitations remain. The non-Newtonian experiment reported here focuses
on low and moderate Weissenberg numbers up to $Wi=0.7$, and the helicity
transfer experiment remains a smooth high-Reynolds-number benchmark rather than
a turbulent flow. Extending the method to higher Weissenberg numbers, turbulent
three-dimensional regimes, adaptive partitions, and
fully parallel local training will require additional stabilization and
systematic hyperparameter studies. The present results nevertheless show that
choosing transfer quantities from the conservation structure of the PDE can
substantially improve the reliability of PINNs for incompressible flow.

%% file: sections/conclusion.tex
\section{Conclusion}
\label{sec:conclusion}

We introduced an invariant-guided PINN framework for
incompressible flow problems. The method trains local PINNs on smaller
partitions, extracts conservative information from the local solutions, uses
that information to initialize or constrain subsequent training, and performs a
global correction to obtain a single full-domain representation. For a
non-Newtonian Oldroyd--B cylinder wake, the method transfers velocity, stress,
and mass flux while avoiding pressure over-constraint at interfaces. For a
Newtonian helicity problem, it trains global spatial PINNs on time slabs,
transfers curl-compatible velocity fields, constrains the kinetic energy, and
then performs a residual global space-time correction while monitoring and
weakly regularizing helicity from the induced vorticity. The
reported experiments demonstrate faster convergence, accurate drag prediction,
small mass-conservation error, and controlled helicity drift.

%% file: sections/declarations.tex
\section*{CRediT authorship contribution statement}

Zheng Lu: Conceptualization, Methodology, Software, Validation,
Visualization, Writing -- original draft. Jiwei Jia: Conceptualization,
Methodology, Supervision, Writing -- review and editing. Bora Aniruddha:
Methodology, Formal analysis, Writing -- review and editing. Xingyu An:
Methodology. Young Ju Lee:
Conceptualization, Supervision, Project administration, Writing -- review and
editing.

\section*{Declaration of competing interest}

The authors declare that they have no known competing financial interests or
personal relationships that could have appeared to influence the work reported
in this paper.

\section*{Data availability}

Data and code will be made available on request.

%% file: sections/acknowledgements.tex
\section*{Acknowledgements}

The authors thank Professor Kaibo Hu for helpful discussions.